\documentclass[aps,prc,reprint,noshowpacs,groupedaddress,onecolumn,floatfix]{revtex4}
\usepackage{amsmath,amssymb}
\usepackage{amsfonts}
\usepackage{amscd} 
\usepackage{graphicx}
\usepackage{color}
\def\beq{\begin{equation}}
\def\eeq{\end{equation}}
\begin{document}
\title{Relativistic invariance in Euclidean
formulations of quantum mechanics.}

\author{Gohin Shaikh Samad}
\affiliation{Department of Mathematics, The University of
Iowa, Iowa City, IA 52242, USA}

\author{W.~N.~Polyzou}
\affiliation{Department of Physics and Astronomy, The University of
Iowa, Iowa City, IA 52242, USA}

\date{\today}

\begin{abstract}

  Relativistic invariance in Euclidean formulations of quantum
  mechanics is discussed.  Relativistic treatments of quantum theory
  are needed to study hadronic systems at sub-hadronic distance
  scales.  Euclidean formulations of relativistic quantum mechanics
  have some computational advantages.  In the Euclidean representation
  the physical Hilbert space inner product is expressed in terms of
  Euclidean space-time variables with no need for any analytic
  continuation.   The identification of the complex Euclidean group
  with the complex Poincar\'e group relates the infinitesimal
  generators of both groups.  In this work explicit representations of
  the Poincar\'e generators in Euclidean space-time variables for all
  positive-mass positive-energy irreducible representations of the
  Poincar\'e group are derived.  The commutation relations are
  checked, both hermiticity and self-adjointness are
  established, and reflection positivity of the kernels is verified.  

\end{abstract}

\thanks{This work supported by the U.S. Department of Energy,
  Office of Science, Grant \#DE-SC16457}

\maketitle

\section{Introduction}

This paper discusses how relativistic invariance is realized in
Euclidean formulations of relativistic quantum theory.  In a quantum
theory relativistic invariance means that quantum observables, which
are probabilities, expectation values and ensemble averages, have the
same value for equivalent experiments that are performed in different
inertial coordinate systems.  This means that experiments performed in an
isolated system cannot be used to distinguish inertial coordinate
systems.  In special relativity different inertial coordinate systems
are related by the subgroup of Poincar\'e group connected to the
identity.  In 1939 Wigner \cite{Wigner:1939cj} showed that a necessary
and sufficient condition for a quantum system to be relativistically
invariant is that vectors representing equivalent quantum states in
different inertial coordinate systems are related by a unitary ray
representation of this subgroup on the Hilbert space of the quantum
theory.

Relativistically invariant quantum theories are needed to study
physics on distance scales that are small enough to be sensitive to
the internal structure of a nucleon.  This is because in order to get
wavelengths short enough to resolve the internal structure of a
nucleon it is necessary to transfer a momentum to the nucleon that is
comparable to or larger than its mass scale.

In quantum theories time evolution is generated by a one-parameter
unitary group.  The infinitesimal generator of this group is the
Hamiltonian, which is a positive self-adjoint operator on the Hilbert
space of the quantum theory.  Because the spectrum of the Hamiltonian,
time can be analytically continued to the lower-half complex time
plane.  For imaginary times, $t \to -i \tau$, the unitary time
evolution group becomes a contractive Hermitian semigroup.  For any
fixed $\tau>0$, $e^{-H\tau}$ has the same eigenvectors as the
Hamiltonian, and the eigenvalues $\lambda$ of $H$ are related to the
eigenvalues $\eta$ of $e^{-H\tau}$ by $\lambda = - \ln (\eta)/\tau$.
This implies that it is possible to solve dynamical problems directly
in a Euclidean representation.  For some applications it is enough to
replace $H$ by $e^{-\tau H}$.  This is a well-behaved bounded operator
with a spectrum on the unit interval $[0,1]$; the parameter $\tau >0$
can be adjusted to be sensitive to different parts of the spectrum of $H$.
Relativistic invariance normally requires an analytic continuation back to real
time.  These observations provide the motivation for investigating
Euclidean approaches to relativistic quantum field theory and quantum
mechanics.

Euclidean approaches were first advocated by Schwinger
\cite{Schwinger:pna}\cite{Schwinger:1959zz} who used the spectral
condition in time-ordered Green's functions to establish the existence
of an analytic continuation to imaginary times.  Independently,
axiomatic treatments of quantum field theory
\cite{Wightman:1980}\cite{jost} led to an understanding of the
analytic properties of vacuum expectation values of products of
fields, also based on the spectral condition.  The Euclidean approach
to quantum field theory was advocated by Symanzik
\cite{Symanzik:1966}\cite{Symanzik:1968zz}, and developed by Nelson
\cite{Nelson:1973}.  Osterwalder and Schrader
\cite{Osterwalder:1973dx}\cite{Osterwalder:1974tc} identified
properties of Euclidean covariant distributions that are sufficient to
reconstruct a relativistic quantum field theory.  Two observations
that are contained in the work of Osterwalder and Schrader are (1)
that an explicit analytic continuation is not necessary to construct a
relativistic quantum theory and (2) the reconstruction of a
relativistic quantum theory is not limited to local field theories.
The discussion that follows is motivated by these two observations.

The Poincar\'e and four-dimensional Euclidean groups are related
because the parameters of both groups can be analytically
continued and the covering group of the resulting complex groups are identical,
$SL(2,\mathbb{C})\times SL(2,\mathbb{C})$.  What this means is that
the real Poincar\'e group can be considered to be a complex subgroup
of the complex Euclidean group, or conversely, the real Euclidean group
can be considered to be a complex subgroup of the complex Poincar\'e group.
These identifications imply formal relations between the infinitesimal
generators of the Poincar\'e group and the real Euclidean group \cite{palle}\cite{olafsson}.
Specifically, if $ P^0_e,\mathbf{P}_e,J^{ij}_e,J^{0i}_e$ satisfy the
commutation relations of the Euclidean Lie Algebra, then the operators
$P^0_m=-i P^0_e,\mathbf{P}_m=\mathbf{P}_e, J^{ij}_m=J^{ij}_e,J^{0i}_m
:= -iJ^{0i}_e$, will satisfy the commutation relations of the
Poincar\'e Lie Algebra.  However, because of the factors of $i$, both
sets of operators cannot be self-adjoint on the same representation of
the Hilbert space.

Osterwalder and Schrader construct a new Hilbert space representation
where the Poincar\'e generators become self adjoint.  Osterwalder and
Schrader start with a representation of a Hilbert space defined with a
Euclidean covariant kernel.  On this space the Euclidean
transformations are norm preserving which defines a unitary
representation of the Euclidean group.  Next they choose an arbitrary
time direction and multiply the final Euclidean time variables in this
kernel by an operator that reverses the sign of all of the final
Euclidean times.  Introducing this time reflection in the Euclidean
kernel breaks the Euclidean invariance and has the effect of making
the Poincar\'e generators constructed from the Euclidean generators
Hermitian on this space.  The integration variables remain unchanged -
they include the Euclidean times.  The problem is that the resulting
quadratic form cannot be positive for all Euclidean test functions.
This is easily seen by taking functions with positive time support and
extending them to be even or odd under time reflection.  Since the
quadratic forms will have opposite signs, they cannot both have positive
norm with this new inner product.  This flaw is fixed by projecting
the test functions on a suitable subspace.  The subspace identified by
Osterwalder and Schrader is the subspace of functions of Euclidean
space-time variables with support for positive absolute and relative
Euclidean times.  The Euclidean kernels are called {\it reflection
positive} if the norms with respect to the inner product with the
Euclidean time reflection is non-negative on this subspace.
Reflection positivity is a constraint on the Euclidean distributions
\cite{Jaffe:2018ftu}. This construction is a specific application of a
general construction based on an abstract notion of reflection
positivity \cite{palle}\cite{olafsson}.

Because this projection is independent of the form of the Euclidean
kernel, cluster properties, which are an important physical
requirement, can be expressed entirely in terms of properties of the
kernel - the range of the projector does not change.  Cluster
properties can be a difficult constraint to satisfy in some representations of
relativistically quantum
mechanics \cite{Sokolov:1977}\cite{Coester:1982vt}\cite{Keister:1991sb},
but it can be easily achieved in the Euclidean approach.

The restriction to positive relative time is because the Euclidean
kernels for irreducible representations of the Poincar\'e group become
singular for zero relative times.  Since identical particles have an
exchange symmetry, this is reflected in the symmetry properties of the
Euclidean kernel.  As long as the relative time supports are disjoint,
the symmetry can be used to reorder the variables so the support
satisfies the positive relative-time condition.  What separates
relativistic quantum theory from local quantum field theory is whether
the symmetries involve all of the coordinates in the kernel or just
separately involve the initial and final coordinates.  This will be
discussed in more detail below.

Reflection positivity is a strong constraint, particularly when it is
combined with Euclidean covariance and cluster properties.  One
consequence is that it implies the spectral condition that Schwinger
originally used to justify the existence of an analytic continuation.
The advantage of the Osterwalder-Schrader reconstruction is that this
analytic continuation is never explicitly needed.

In this paper a Euclidean relativistic theory is defined by a
finite or infinite collection of Euclidean covariant tempered
distributions
\beq
S_{m:n} (x_m, \cdots ,x_1 ; y_1, \cdots, y_n). 
\label{a.1}
\eeq
These kernels contain the dynamics.
The kernels satisfy the permutation symmetry,
\[
S_{m:n} (x_m, \cdots ,x_1 ; y_1, \cdots, y_n) =
(\pm)^{\vert \sigma \vert}
 S_{m:n} (x_{\sigma (m)} \cdots ,x_{\sigma (1)} ; y_1, \cdots, y_n) 
\]
\beq
= (\pm)^{\vert \sigma \vert}  
S_{m:n} (x_m, \cdots ,x_1 ; y_{\sigma (1)}, \cdots, y_{\sigma (n)} ) 
\label{a.2}
\eeq
where $\sigma()$ is a permutation on $m$ or $n$ objects,
$\vert \sigma \vert$ is 0 if $\sigma$ is an even permutation and 1 if
it is an odd permutation. The $+$ sign is for Bosons and the $(-)$
sign is for Fermions.  For local quantum field theories the collection
must be infinite, $S_{m:n} = S_{k:l}$ whenever $m+n=k+l$, and the
permutation symmetry is with respect to all $n+m$ variables.  In
(\ref{a.1}-\ref{a.2}) the $x_n$ can also include spin degrees of
freedom.

The symmetry in the local field theory case arises because the domain
of analyticity, that comes from the spectral condition, can be
extended by complex Lorentz transformations.  The extended domain of
analyticity includes real space-like separated points (Jost points)
that allow the local
fields to be reordered \cite{Wightman:1980}, relating Green's functions with
permuted arguments.  This symmetry is not assumed in this work.  One
consequence of relaxing this condition is that it is possible to have
different $N$-point Green's functions for different numbers of initial
and final coordinates.

The setting for a quantum theory is a Hilbert space.  A dense set of
vectors in the Euclidean representation of the Hilbert space are
sequences of Schwartz test functions of Euclidean space-time variables
\beq
\{\psi_n (x_1 \cdots x_n)\}_{n=0}^N
\label{a.3}
\eeq
that vanish unless the Euclidean times satisfy
$0 < x_1^0 < x_1^0 < \cdots < x_n^0$.

The Hilbert space inner product is
\[
\langle \psi \vert \phi \rangle =
\sum_{mn}
\int d^4x_m \cdots d^4x_1 d^4y_1\cdots  d^4 y_n
\psi_m^* (\theta x_m, \cdots , \theta x_1) \times
\]
\beq
S_{m:n} (x_m, \cdots ,x_1 ; y_1, \cdots, y_n)
\phi_n  (y_1, \cdots , y_n)
\label{a.4}
\eeq
where $\theta$ represents Euclidean time reflection,
$\theta (x_i^0,\mathbf{x}_i) := (-x_i^0,\mathbf{x}_i)$ .
Because of the
assumed symmetry properties of the Euclidean distributions, as long as
the Euclidean time supports in the functions are ordered for one set
of Euclidean times, the permutation symmetry can be used to replace
them for one that is ordered as above.

One property of this representation of the Hilbert space, where the
inner products has a non-trivial kernel, is that distributions like
delta functions represent normalizable vectors.    

For this to be a Hilbert space scalar product, this quantity must be
non-negative whenever $\{ \psi_m \} = \{\phi_n \}$.  This condition is
called reflection positivity.  In general there can be $0$-norm
vectors.  The Hilbert space vectors are Cauchy sequences of
equivalence classes of vectors, where two vectors are in the same
class if the norm of their difference vanishes.  This distinction will be
ignored in what follows.
For free particles, reflection positivity restricts the form of the
allowed distributions
\cite{Widder:1931}\cite{Widder:1934}\cite{Widder:1941}.  They are
singular when the relative Euclidean coordinates vanish.  The restriction picks a
domain where the scalar products are finite.

Because the Euclidean time reflection breaks the Euclidean invariance,
both Euclidean time translation and rotations in Euclidean space-time
planes are no longer unitary on this space.  These transformations are
nevertheless defined on this space with restricted domains; they
represent translations in imaginary time and boosts with imaginary
rapidity.  The infinitesimal forms of these elementary Euclidean
transformations can be used to construct both the Hamiltonian and
Lorentz boost generators.

The purpose of this work is to give a detailed discussion of how
relativistic invariance is realized in these theories.  Rather than
consider a general set of Euclidean covariant kernels,
this work is limited to Euclidean representations of irreducible
representation of the Poincar\'e group \cite{Polyzou:2019a}.  There are
two motivations for this.  The first is that the kernels for these
representations are known, so it is possible to understand domain
issues related to the properties of the kernel and give explicit
representations for the Poincar\'e generators.
The second motivation
is that any unitary representation of the Poincar\'e group can be
decomposed into a direct integral of irreducible representations.  In
a relativistic quantum theory these can be identified with the
complete set of one-body states plus multi-particle in or out
scattering states.  These states either transform irreducibly or as
tensor products of irreducible representations.  The kernel of a
general interacting model should be related to the direct integral of
irreducible kernels by a unitary transformation.  The construction of
this direct integral from a general set of Euclidean covariant
distributions is the relativistic analog of diagonalizing the
Hamiltonian in non-relativistic quantum mechanics.  This will not be
considered in this work.

In the next section the Poincar\'e group and its relation to the
Euclidean group is discussed.  In section three unitary
representations of the Poincar\'e group are discussed, along with
structure of positive mass irreducible representations.  Section 4
contains explicit forms of Euclidean covariant kernels of irreducible
representations of the Poincar\'e group for any mass and spin.  They
are shown to be reflection positive.  Explicit forms for all the
Poincar\'e generators are constructed, commutation relations are
verified, and the generators are shown to be symmetric with respect
the inner product with the Euclidean time reflection.  Section 5
discuss the self-adjointness of the Hamiltonian and rotationless boost
generators.  Section 6 has a brief discussion of finite Poincar\'e
transformations.  The results are summarized in section 7.

\section{Background}

The Poincar\'e group is the group of space-time transformations that
relate different inertial reference frames in the theory of special relativity.
It is the symmetry group that preserves the proper time $\tau_{ab}$,
or proper distance, $d_{ab}$, between any two events with space-time
coordinates $x_a^{\mu} , x_b^{\mu}$
\beq
-\tau_{ab}^2=d_{ab}^2=\eta_{\mu\nu}(x_a-x_b)^{\mu}(x_a-x_b)^{\nu}, 
\label{b.1}
\eeq
where $\eta_{11}=\eta_{22}=\eta_{33}=-\eta_{00}=1$, $\eta_{\mu\nu}=0$ for
$\mu \not=\nu$ is the Minkowski metric tensor.  Repeated indices
are assumed to be summed.
The most general point transformation,
$x'^\mu=f^\mu(x)$ satisfying (\ref{b.1}) has the form
\beq
x^{\mu}\to x'^{\mu}=\Lambda^{\mu}{}_{\nu} x^\nu+a^\mu 
\label{b.2}
\eeq
where $\Lambda^{\mu}{}_{\nu}$ is a Lorentz transformation satisfying 
\[
\eta_{\mu\nu}=\Lambda^{\alpha}{}_\mu\eta_{\alpha\beta} \Lambda^{\beta}{}_\nu
\]
or in matrix form 
\beq
\eta=\Lambda^t\eta\Lambda .
\label{b.3}
\eeq
Equations (\ref{b.2}) and (\ref{b.3}) are relativistic generalizations of the
fundamental theorem of rigid body motion, which asserts that any motion
that preserves the distance between points in a rigid-body in a
composition of an orthogonal transformation and a translation.

The full Poincar\'e group contains discrete transformations that are
not associated with special relativity.  
Equation (\ref{b.3}) implies that
\beq
\mbox{det}(\Lambda)^2=1 \qquad \mbox{and} \qquad (\Lambda_0^0)^2=
1+\sum_{i}(\Lambda_i^0)^2 .
\label{b.4}
\eeq
This means that the Lorentz group can be decomposed
into four topologically disconnected components
\begin{itemize}
\item 
$\mbox{det}(\Lambda)=1, \qquad(\Lambda_0^0) \geq 1$; includes identity
\item
$\mbox{det}(\Lambda)=-1, \qquad(\Lambda_0^0) \geq 1$; includes space reflection
\item
$\mbox{det}(\Lambda)=-1, \qquad (\Lambda_0^0) \leq-1 $; includes time reversal
\item
$\mbox{det}(\Lambda)=1, \qquad(\Lambda_0^0) \leq -1$; includes space-time reversal .
\end{itemize}
Since the discrete symmetries of space reflection and time reversal
are not symmetries of the weak
interaction, the symmetry group associated with special relativity is
normally considered to be the subgroup of Poincar\'e transformations
that is continuously connected to the identity. This subgroup contains
the active transformations that can be experimentally realized.

The relation between the Lorentz group and the
four-dimensional orthogonal group can be understood by
expressing Minkowski, $x^{\mu}$, and Euclidean, $x_e^{\mu}$, four
vectors as $2\times 2$ matrices:
\beq
X_m=x^\mu\sigma_\mu=
\begin{pmatrix}
x^0+x^3 &x^1-ix^2  \\
x^1+ix^2 &x^0-x^3 
\end{pmatrix}\; \qquad  x^\mu=\frac{1}{2}\mbox{Tr}(X\sigma_\mu)
\label{b.5}
\eeq
\beq
X_e=x_e^\mu\sigma_{e\mu}=
\begin{pmatrix}
ix_e^0+x^3 &x^1-ix^2  \\
x^1+ix^2 &ix_e^0-x^3 
\end{pmatrix}\; \qquad  x_e^\mu=\frac{1}{2}\mbox{Tr}(X_e\sigma^{\dagger}_{e\mu}).
\label{b.6}
\eeq
where $\sigma_i=\sigma_{ei}$ are the Pauli matrices, $\sigma_0$ is the identity and
$\sigma_{e0}= i \sigma_0$.  The determinants of these matrices are
related to the Minkowski and Euclidean line elements respectively:
\beq
\mbox{det}(X_m) = (x^0)^2 - \mathbf{x}\cdot \mathbf{x}
\qquad 
\mbox{det}(X_e) = - \left ( (x_e^0)^2 + \mathbf{x}\cdot \mathbf{x}
\right ).
\label{b.7}
\eeq
The linear transformations that preserve the determinant and hermiticity of
$X_m$ have the form
\beq
X_m \to X_m' = \pm A X_m A^{\dagger}
\qquad \mbox{det} (A)=1.
\label{b.8}
\eeq
The (-) sign represents a space-time reflection, which is not
considered part of the symmetry group of special relativity.  The
group of complex $2 \times 2$ matrices with $\mbox{det}(A)=1$ is
$SL(2,\mathbb{C})$.  Similarly linear transformations
corresponding to real four-dimensional orthogonal transformations have
the general form
\beq
X_e \to X_e' =  A X_e B^t \qquad A,B \in SU(2).
\label{b.9}
\eeq
Transformations of the form
\beq
X_e \to X_e' =  A X_e B^t \qquad 
X_m \to X_m' =  A X_m B^t
\label{b.10}
\eeq
with both $A$ and $B$ in $SL(2,\mathbb{C})$ preserve both the
Minkowski and Euclidean line elements.  However they do not
preserve the reality of the four vectors.  They represent complex
Lorentz or orthogonal transformations.

This shows that the covering group of both the complex Lorentz and
complex orthogonal group is
$SL(2,\mathbb{C}) \times SL(2,\mathbb{C})$.  This means that the real
Lorentz group can be considered to be a subgroup of the complex
orthogonal group; similarly the real orthogonal group can be
considered to be a complex subgroup of the Poincar\'e group.  The
relevant relation that will be exploited in this work is that
Euclidean rotations that involve a space and the Euclidean time
coordinate can be identified with Lorentz boosts with complex
rapidity.

For the full Poincar\'e group it is necessary to include translations.
Euclidean time translations by $\tau$ are identified with Minkowski
time translations with $t=-i\tau$. 

\section{Unitary representations of the Poincar\'e group}

In this section Poincar\'e group elements are labeled by $(\Lambda,A)$
where $\Lambda$ is a $SL(2,\mathbb{C})$ matrix and $A$ is a $2 \times 2$
Hermitian matrix representing a translation.  In this representation
Poincar\'e transformations have the form
\beq
X' = \Lambda X \Lambda^{\dagger} +A
\label{c.1}
\eeq
where the group multiplication law is
\beq
(\Lambda_2 , A_2) (\Lambda_1 , A_1)= (\Lambda_2 \Lambda_1 , \Lambda_2
A_1 \Lambda_2^{\dagger} + A_2).  
\label{c.2}\eeq
Four vector representations of these equations are
\beq
x^{\mu \prime} = \Lambda^{\mu}{}_{\nu} x^{\nu} + a^{\mu}
\label{c.3}
\eeq
\beq
(\Lambda^{\mu}{}_{\nu},a^{\mu} ) =
(\Lambda_2^{\mu}{}_{\alpha} \Lambda_1^{\alpha}{}_{\nu},
\Lambda_2^{\mu}{}_{\alpha}a_1^{\alpha} + a_2^{\mu}),
\label{c.4}
\eeq
where the
four vector and $2\times 2$ representations are related by 
\beq
a^{\mu}:= {1 \over 2} \mbox{Tr}(\sigma_{\mu}A) 
\qquad 
\Lambda^{\mu}{}_{\nu} := {1 \over 2} \mbox{Tr}(\sigma_{\mu}\Lambda \sigma_{\nu} \Lambda^{\dagger}).
\label{c.5}
\eeq
$SL(2,\mathbb{C})$ is a six parameter group.  It has six independent
one-parameter subgroups
\beq
\Lambda_r (\pmb{\theta})= e^{{i \over 2 }\pmb{\sigma}\cdot \pmb{\theta}}
\qquad
\Lambda_b (\pmb{\rho}) = e^{{1 \over 2 }\pmb{\sigma}\cdot \pmb{\rho}}
\label{c.6}
\eeq
corresponding to rotations about three different axes and rotationless
Lorentz boosts in three different directions.  In these expressions
$\pmb{\theta}$ represents the angle and axis of a rotation while
$\pmb{\rho}$ represents the rapidity and direction of a rotationless boost.
The polar decomposition expresses a general $SL(2,\mathbb{C})$ matrix $\Lambda$
as a product of a rotation ($\Lambda_r$ unitary) followed by
rotationless boost ($\Lambda_b$ positive Hermitian):
\beq
\Lambda = \Lambda_{b}\Lambda_{r} 
\label{c.7}
\eeq
where
\beq
\Lambda_{b} :=
(\Lambda \Lambda^{\dagger})^{1/2} = \Lambda_b (\pmb{\rho})
\qquad 
\Lambda_{r} :=
(\Lambda \Lambda^{\dagger})^{-1/2}\Lambda =\Lambda_r (\pmb{\theta}).
\label{c.8}
\eeq
A unitary representation of the Poincar\'e group (inhomogeneous
$SL(2,\mathbb{C})$) is a set of unitary operators
$U(\Lambda ,A)$, labeled by elements of $SL(2,\mathbb{C}$)
satisfying
\beq
U(\Lambda_2,A_2) U(\Lambda_1,A_1)=
U(\Lambda_2 \Lambda_1,\Lambda_2 A_1 \Lambda_2^{\dagger} +A_2)
\label{c.9}
\eeq
\beq
U(I,0) =I
\label{c.10}
\eeq
\beq
U^{\dagger}(\Lambda ,A) = U^{-1}(\Lambda ,A)=
U (\Lambda^{-1} ,-\Lambda^{-1} A (\Lambda^{\dagger})^{-1} ) .
\label{c.11}
\eeq
The Poincar\'e group is a 10 parameter group.  Infinitesimal generators
are the 10 self-adjoint operators defined by
\beq
H= i {d \over da^0} U(I,a^0\sigma_0)_{\vert_{a_0=0}}
\label{c.12}
\eeq
\beq
P^i= - i {d \over da^j} U(I,a^j\sigma_j)_{\vert_{a_i=0}}
\label{c.13}
\eeq
\beq
J^i= - i {d \over d\theta} U(e^{i{\theta\over 2} \sigma_j},0)_{\vert_{\theta=0}}
\label{c.14}
\eeq
\beq
K^i= - i {d \over d\rho} U(e^{{\rho\over 2} \sigma_j},0)_{\vert_{\rho=0}}
\label{c.15}
\eeq
where there is no sum in (\ref{c.13}) over the repeated $j$, and
$j\in \{1,2,3\}$ in (\ref{c.13}-\ref{c.15}).
The group representation property (\ref{c.9}) implies that these generators
satisfy commutation relations
\beq
[J^i ,J^j] = i \epsilon_{ijk} J^k
\qquad
[J^i ,P^j] = i \epsilon_{ijk} P^k
\qquad
[J^i ,K^j] = i \epsilon_{ijk} K^k
\label{c.16}
\eeq
\beq
[K^i ,K^j] = -i \epsilon_{ijk} J^k
\qquad
[J^i ,H] = 0
\qquad
[P^i,H] = 0
\label{c.17}
\eeq
\beq
[K^j,H] =i P^j
\qquad
[K^i ,P^j] = i \delta_{ij}H .
\label{c.18}
\eeq

These operators are components of a four vector, $P^{\mu}$, and an anti-symmetric
tensor operator, $J^{\mu \nu}$,
\beq
P^{\mu}  = (H,\mathbf{P})
\qquad
J^{\mu \nu} =
\left (
\begin{array}{cccc}
0 & -K^1 & -K^2 & -K^3 \\
K^1 & 0 & J^3 & -J^2 \\
K^2 & -J^3 & 0 & J^1 \\
K^3 & J^2 & -J_1 & 0\\ 
\end{array}
\right ) .
\label{c.19}
\eeq
There are two independent polynomial invariants
\beq
M^2 = (P^0)^2 -\mathbf{P}^2 = - P^{\mu}P_{\mu} 
\label{c.20}
\eeq
and
\beq
W^2 = W^{\mu}W_{\mu}
\qquad
W^{\mu} = -{1 \over 2} \epsilon^{\mu \nu \alpha \beta} P_{\nu}J_{\alpha \beta}.
\label{c.21}
\eeq
where $W^{\mu}$ is called the Pauli-Lubanski vector.
When $M\not=0$ the spin is defined by
\beq
S^2 = W^2/M^2 .
\label{c.22}
\eeq
A spin vector can be defined by an {\it operator}
rotationless (canonical) 
boost that
transforms the angular momentum tensor to the rest frame:
\beq
s^i = \epsilon_{ijk} 
\Lambda_c^{-1} (P)^{j}{}_{\mu}
\Lambda_c^{-1} (P)^{k}{}_{\nu}J^{\mu \nu} 
\label{c.23}
\eeq
where
\beq
\Lambda_c (P)^{\mu}{}_{\nu} =
\left (
\begin{array}{cc}
V^0 & \mathbf{V} \\
\mathbf{V} & I + {\mathbf{V}\otimes \mathbf{V} \over 1+ V^0} \\
\end{array}  
\right )
\qquad V^{\mu}= P^{\mu}/M
\label{c.24}
\eeq
and $P^{\mu}$ and $M$ are
considered operators.  Note that $\Lambda_c (p) = \Lambda_b (\pmb{\rho})$
with
\beq
\mathbf{V} = \mathbf{P}/M =  \hat{\pmb{\rho}} \sinh (\rho).
\eeq
This spin vector is called the canonical spin; other types
of spin vectors (helicity, light-front spin) are
related to the canonical spin by momentum
dependent rotations.  For the purpose of this work it is sufficient to
consider the canonical spin.
The canonical spin can also be expressed in terms of the Pauli Lubanski vector:
\beq
\left
(  
\begin{array}{c}
0 \\
\mathbf{s}_c \\
\end{array}
\right ) =
- {1 \over 2M } \Lambda_c^{-1} (P)^{\mu}{}_{\nu} W^{\mu}. 
\label{c.25}
\eeq
The components of the spin satisfy $SU(2)$ commutations relations:
\beq
[s_i,s_j] = i \epsilon_{ijk} s^k .
\label{c.27}
\eeq
With these definitions, for $M>0$, $M^2,s^2,\mathbf{P},s_z$ are a
maximal set of commuting self-adjoint functions of the Poincar\'e
generators.  The spectrum of each component of $\mathbf{P}$ is the
real line since each component of $\mathbf{P}$ can be boosted to any
value.  Similarly the spectrum of spins are restricted to be integral
or half integral as a consequence of the $SU(2)$ commutations
relations.  In a general system these commuting observables are not
complete; they can be supplemented by additional Poincar\'e-invariant
degeneracy quantum numbers, which will be denoted by $\alpha$.  A
basis for the Hilbert space are the simultaneous eigenstates of
$M,S^2,\alpha,\mathbf{P},s_z $,
\beq
\vert (m,s,\alpha)
\mathbf{p},\mu \rangle .
\label{c.28}
\eeq
Because these vectors are constructed out of eigenvalues of functions of
$P^{\mu}$ and $J^{\mu \nu}$, which have well-defined the Poincar\'e
transformation properties,  the Poincar\'e transformation properties
of these basis state follow from the definitions
\beq
U(\Lambda,a)\vert(m,s,\alpha) p,\mu \rangle =e^{i\Lambda p\cdot a}\vert
(m,s,\alpha)\Lambda p,\nu
\rangle
D_{\nu\mu}^{j}[R_{ws}(\Lambda,p)] \sqrt{\frac{\omega_{m}(\Lambda p)}{\omega_{m}(p)}}
\label{c.29}
\eeq
where, $R_{cw}(\Lambda,p):=\Lambda_c^{-1}(\Lambda p)\Lambda \Lambda_c(p)$ is the
canonical-spin Wigner rotation, $\Lambda_c(p)= e^{{1\over 2}\pmb{\rho}\cdot \pmb{\sigma}}$
where $\pmb{\rho}$ is the rapidity of a particle of mass $m$ and momentum
$\mathbf{p}$, 
and $\omega_{m}(p):=\sqrt{m^2+\mathbf{p}^2}$ is the energy of the particle.

The Wigner $D$-function is the finite dimensional unitary representation of the
rotation group in the $\vert s,\mu\rangle $ basis \cite{rose}:
\[
D_{m,m'}^{s}[R]=\langle s,\mu\vert U(R) \vert s,\mu'\rangle  =
\]
\[
\sum_{k=0}^{s+\mu}\frac{\sqrt{(s+\mu)!(s+\mu')!(s-\mu)!(s-\mu')!}}{k!
  (s+\mu'-k)!(s+\mu-k)!(k-\mu-\mu')!}R_{++}^{k}R_{+-}^{s+\mu'-k}
R_{-+}^{s+\mu-k}R_{--}^{k-\mu-\mu'}
\]
where
\beq
R=
\left (
\begin{array}{cc}
R_{++} & R_{+-}\\
R_{-+} & R_{--}
\end{array}
\right ) = e^{{i\over 2}\pmb{\theta}\cdot \pmb{\sigma}}=
\sigma_0 \cos ({\theta \over 2})  + i \hat{\pmb{\theta}}\cdot \pmb{\sigma}
\sin ({\theta \over 2})
\label{c.30}
\eeq
is a $SU(2)$ matrix.  Because $D^s_{\mu \nu}[R]$ is a degree $2s$
polynomial in the matrix elements of $R$, and
$R=e^{i {\pmb{\theta}\cdot \pmb{\sigma}\over 2}}$ is an entire function of the
angles, $\pmb{\theta}$,  it follows that
$D_{\mu,\mu'}^{s}[e^{i {\pmb{\theta}\cdot \pmb{\sigma}\over 2}}]$ is
an entire function of all three components of $\pmb{\theta}$.  This means
that the group representation property
\beq
\sum_{\mu''} D_{\mu,\mu''}^{s}[R_2] D_{\mu'',\mu'}^{s}[R_1] -
D_{\mu,\mu'}^{s}[R_2R_1]=0, 
\label{c.31}
\eeq
and the formulas for adding angular momenta
\beq
D_{\mu,\mu'}^{s}[R] -
\sum_{ \mu_1 \mu_2  \mu_1' \mu_2'}
\langle s,\mu\vert s_1, \mu_1, s_2, \mu_2 \rangle 
D_{\mu_1,\mu_1'}^{s_1}[R] D_{\mu_2,\mu_2'}^{s_2}[R]
\langle s_1, \mu_1', s_2', \mu_2' \vert s , \mu'\rangle =0
\label{c.32}
\eeq
and
\beq
D_{\mu_1,\mu_1'}^{s_1}[R] D_{\mu_2,\mu_2'}^{s_2}[R] -
\sum_{s \mu \mu'} \langle s_1, \mu_1, s_2, \mu_2 \vert s , \mu\rangle
D_{\mu,\mu'}^{s}[R] 
\langle s,\mu'\vert s_1, \mu_1', s_2, \mu_2' \rangle =0,
\label{c.33}
\eeq
which hold for real angles, can be analytically continued to complex
angles.  This means that (\ref{c.31}-\ref{c.33}) also hold when the
$SU(2)$ matrices $R$ are
replaced by $SL(2,\mathbb{C})$ matrices.
In these expressions, $\langle s,\mu\vert s_1, \mu_1, s_2,
\mu_2 \rangle$, are $SU(2)$ Clebsch-Gordan coefficients.  While the
analytic continuation preserves the group representation and angular
momentum addition properties, it does not preserve unitarity.

\section{Euclidean Formulation}

The common property of any relativistic quantum theory is that it can
be decomposed into a direct integral of irreducible representations.
The structure of irreducible representations of the Poincar\'e group
in the Euclidean representation can be understood by starting with
Minkowski-space irreducible representations of the Poincar\'e group.
This work considers only positive-mass positive-energy
representations.  These can be expressed in a basis of simultaneous
eigenstates of the mass, spin, linear momentum and $z$-component of
the canonical spin.  The action of the unitary representation of the
Poincar\'e group on this basis is given by (\ref{c.29}).
This is unitary for basis vectors with the normalization:
\beq
\langle (m',s')p',\mu'
\vert (m,s) p,\mu \rangle = \delta_{m'm}\delta_{s's} \delta
(\mathbf{p}'-\mathbf{p}) \delta_{\mu' \mu}.
\label{d.1}
\eeq
Because of the unitarity of $R_{wc}(\Lambda,p)$, the $SU(2)$
Wigner rotation can be
expressed in two equivalent ways:
\beq
R_{wc}(\Lambda,p) =
\Lambda_c^{-1} (\Lambda p) \Lambda \Lambda_c (p) = \Lambda_c^{\dagger}
(\Lambda p) (\Lambda^{\dagger})^{-1} \Lambda_c^{\dagger -1} (p) .
\label{d.2}
\eeq
The $SL(2,\mathbb{C})$ group representation property (\ref{c.31})
implies that the unitary representation of the Wigner rotation can be
factored into a product of three finite-dimensional representations of
$SL(2,\mathbb{C})$ in two different ways:
\beq
D_{\nu\mu}^{s}[R_{wc}(\Lambda,p)]=
\sum_{\alpha \beta} D_{\nu\alpha}^{s}[\Lambda_c^{-1} (\Lambda p)]
D_{\alpha\beta}^{s}[\Lambda] D_{\beta\mu}^{s}[\Lambda_c (p)]
\label{d.3}
\eeq
or
\beq
D_{\nu\mu}^{s}[R_{wc}(\Lambda,p)]= \sum_{\alpha \beta}
D_{\nu\alpha}^{s}[\Lambda_c^{\dagger} (\Lambda p)]
D_{\alpha\beta}^{s}[(\Lambda^{\dagger})^{-1}]
D_{\beta\mu}^{s}[(\Lambda_c^{\dagger})^{-1} (p)] .
\label{d.4}
\eeq
These relations can be used to rewrite equation (\ref{c.29}) in
terms of new Lorentz covariant basis states:
\[
U(\Lambda,a) \underbrace{ \sum_\alpha
\vert (m,j) p,\alpha \rangle
D_{\alpha\mu}^{s}[\Lambda_b^{-1} (p)] \sqrt{\omega_m(p)}}_
{\vert (m,j) p,\mu \rangle_{cov}}
=
\]
\beq
e^{i\Lambda p\cdot a}\sum_\beta
\underbrace{\sum_\alpha \vert (m,j)\Lambda p,\alpha \rangle 
D_{\alpha\beta}^{s}[\Lambda_c^{-1} (\Lambda p)]
\sqrt{\omega_{m}(\Lambda p)}}_
{\vert (m,j) \Lambda p,\beta \rangle_{cov}}
D_{\beta\mu}^{s}[\Lambda]
\label{d.5}
\eeq
or 
\[
U(\Lambda,a) \underbrace{\sum_\alpha
\vert (m,j) p,\alpha \rangle
D_{\alpha\mu}^{s}[\Lambda_c^{-\dagger} (p)] \sqrt{\omega_m(p)}}_
{\vert (m,j) p,\mu \rangle_{cov*}}
=
\]
\beq
e^{i\Lambda p\cdot a}\sum_\beta
\underbrace{\sum_\alpha \vert (m,j)\Lambda p,\alpha \rangle
D_{\alpha\beta}^{s}[\Lambda_c^{\dagger} (\Lambda p)]
\sqrt{\omega_{m}(\Lambda p)}}_
{\vert (m,j) \Lambda p,\beta \rangle_{cov*}}
D_{\beta\mu}^{s}[(\Lambda^{\dagger})^{-1}].
\label{d.6}
\eeq
These expressions replace the states (\ref{c.28}) that transform
covariantly with respect to the Poincar\'e group with states that
transform covariantly with respect to $SL(2,\mathbb{C})$.  The
transformations relating the Lorentz and Poincar\'e covariant
representations are
invertible, however there are two distinct Lorentz covariant
representations, because while $R=(R^{\dagger})^{-1}$ for $R\in
SU(2)$, the corresponding representations in $SL(2,\mathbb{C})$ are
inequivalent.  These two representations are called right and left
handed representations for reasons that will become apparent.

In the Lorentz covariant representations, (\ref{d.5}) and (\ref{d.6}),
this equivalence can be used to show that the Hilbert space inner
product of two $SL(2,\mathbb{C})$ covariant wave functions has a
non-trivial kernel
\[
\langle \psi \vert \phi \rangle  =
\sum_\mu \int
\langle \psi \vert (m,j) p,\mu \rangle
d\mathbf{p} \langle (m,j) p,\mu \vert \phi \rangle=
\]
\beq
\int \sum_{\mu \nu} \langle \psi \vert (m,j) p,\mu \rangle_{cov}{}
D_{\mu\nu}^{j}[p\cdot\sigma]
2 \delta (p^2+m^2)\theta (p^0) d^4 p{}
_{cov}\langle (m,j) p,\nu \vert \phi \rangle
\label{d.7}
\eeq
\[  
\langle \psi \vert \phi \rangle  =
\int \sum_\mu
\langle \psi \vert (m,j) p,\mu \rangle
d\mathbf{p} \langle (m,j) p,\mu \vert \phi \rangle=
\]
\beq
\int \sum_{\mu \nu}  \langle \psi \vert (m,j) p,\mu \rangle_{cov*}{}
D_{\mu\nu}^{j}[\Pi p\cdot\sigma]
2 \delta (p^2+m^2)\theta (p^0) d^4p{}
_{cov*}\langle (m,j) p,\nu \vert \phi \rangle
\label{d.8}
\eeq
where $\Lambda_c (p) \Lambda^{\dagger}_c (p)= {\sigma}\cdot {p}$ and
$\Lambda^{-1}_c (p) (\Lambda^{\dagger}_c)^{-1}(p) = \Pi {p} \cdot {\sigma}$,
was used in these equations.
$\Pi$ is the space reflection
operator and $p\cdot\sigma = \omega_m(p)\sigma_0 + \mathbf{p}\cdot
\pmb{\sigma}$.   These equations explain why
(\ref{d.7}) and (\ref{d.8})  are called right and left handed
representations.  These kernels are, up to normalization,
spin-$s$ two-point Wightman functions \cite{Wightman:1980}.

The motivation for considering these $SL(2,\mathbb{C})$ covariant
representations is that they are naturally related to the
corresponding Euclidean covariant representations. 

To show this let $f(x_e,\mu)$ and $g(y_e,\nu)$ be functions of
Euclidean space-time variables and spins with positive Euclidean-time support.
Consider the following Euclidean covariant kernel:
\beq
S^s_{e}(x_e,\mu;y_e,\nu):=
\int d^4p
{2 \over (2 \pi)^4} {e^{i p_e \cdot (x_e -y_e)}
\over p_e^2 + m^2}D^s_{\mu \nu}(p_e \cdot \sigma_e) .
\label{d.9}
\eeq
The physical Hilbert space inner product (\ref{a.4}) for
this Euclidean Green's function
has the form
\[
\int\sum_{\mu \nu} d^4x_ed^4y_e f^*(\theta x_e,\mu)
S^s_{e}(x_e,\mu;y_e,\nu)g(y_e,\nu) =
\]
\[
\int\sum_{\mu \nu} d^4p_e f^*(\theta x_e,\mu)
{2 \over (2 \pi)^4} {e^{i p_e \cdot (x_e -y_e)}
\over p_e^2 + m^2}D^s_{\mu \nu}(p_e \cdot \sigma_e)
g(y_e,\nu) =
\]
\beq
\int\sum_{\mu\nu} \psi^* (\mathbf{p},\mu) {d \mathbf{p} \over \omega_m (\mathbf{p})} 
D^s_{\mu \nu}(p \cdot \sigma) \phi(\mathbf{p},\nu)
\label{d.10}
\eeq
where 
\beq
\psi^*(\mathbf{p},\mu):=   
{1 \over (2 \pi)^{3/2}} \int d\mathbf{x}d\tau e^{i \mathbf{p} \cdot \mathbf{x} - \omega_m (\mathbf{p}) \tau} f^*(\mathbf{x},\tau,\mu) 
\label{d.11}
\eeq
and 
\beq
\phi(\mathbf{p},\nu):= 
{1 \over (2 \pi)^{3/2}} \int d\mathbf{x}d\tau  e^{-i \mathbf{p} \cdot \mathbf{x} - \omega_m (\mathbf{p}) \tau} g(\mathbf{x},\tau,\nu).
\label{d.12}
\eeq
The Euclidean time-support condition ensures that the Laplace
transforms with respect to the Euclidean times in (\ref{d.11}) and
(\ref{d.12}) are well defined.  The resulting kernel in (\ref{d.10})
is identical to the covariant kernel in (\ref{d.7}) after performing
the integrals over the $p^0_e$.

This shows that the ``Euclidean'' inner product
(\ref{d.10}) can be identified with the corresponding Lorentz covariant
inner product, which itself is identical to the original Poincar\'e
covariant inner product. 

This means that 
\beq
S^s_r(x_e, \mu ; y_e, \nu):= \int
{2 d^4p \over (2 \pi)^4} {e^{i p_e \cdot (x_e -y_e)}
\over p_e^2 + m^2}D^s_{\mu \nu}(p_e \cdot \sigma_e)
\label{d.13}
\eeq
is a Euclidean covariant reflection positive kernel for right
handed representations of mass $m$ and spin $s$ respectively.  The corresponding
kernel for left-handed representations is 
\beq
S^s_l(x_e, \mu ; y_e, \nu):= \int 
{2 d^4 p\over (2 \pi)^4} {e^{i p_e \cdot (x_e -y_e)}
\over p_e^2 + m^2}D^s_{\mu \nu}(\Pi p_e \cdot \sigma_e).
\label{d.14}
\eeq
Space reflection interchanges right and left-handed representations.
The space reflection operator does not commute with the Euclidean
covariant kernel.  This implies that space reflected states will not
transform correctly under Lorentz transformations in these Lorentz
covariant representations.  Kernels for systems that allow a linear
representation of space reflection can be constructed by taking direct
sums of right and left handed kernels. 

The kernels (\ref{d.13}-\ref{d.14}) can be evaluated analytically
using the methods in
\cite{bogoliubov}.
The results are
\[
S^s_{r} (z_e, \mu , \nu) := 
{2 \over (2\pi)^4} \int {d^4p \over p_e^2 + m^2}
D^s_{\mu \nu}(p_e \cdot \sigma_e)
e^{i p \cdot z_e} =
\]
\beq
D^s_{\mu \nu}(-i \nabla_{ze} \cdot \sigma_e)
{2m^2
\over (2 \pi)^2}{K_1 (m\sqrt{z_0^2 +\mathbf{z}^2})
\over m\sqrt{z_0^2 +\mathbf{z}^2}}
\label{d.15}
\eeq
\[
S^s_{l} (z_e ,\mu , \nu) := 
{2 \over (2\pi)^2} \int {d^4p  \over p_e^2 + m^2}
D^s_{\mu \nu}(\Pi p_e \cdot \sigma_e)e^{i p_e \cdot z_e} =
\]
\beq
D^s_{\mu \nu}(-i \Pi  \nabla_{ze} \cdot \sigma_e)
{2m^2
\over (2 \pi)^2}{K_1 (m\sqrt{z_0^2 +\mathbf{z}^2})
\over m\sqrt{z_0^2 +\mathbf{z}^2}}
\label{d.16}
\eeq
where $z_e = x_e-y_e$.
Note that ${K_1 (\eta)
\over \eta}$ behaves like $1/\eta^2$ near the origin.   
Since 
$D^s_{\mu \nu}(-i \nabla_{ze} \cdot \sigma_e )$
is a degree $2s$ polynomial in $-i \nabla_{ze}$, these kernels
have power law singularities at the origin, but fall off exponentially
for large values of $z_e^2$, The restriction of the support of the
vectors to positive Euclidean time ensures that $z_e^2 >0$, so the
singularity at $z_e=0$ never causes a problem.  These Green's
functions are reflection positive on this space.
This is because $D^s_{\mu \nu}(p\cdot \sigma)
$ factors into a product of a matrix and its adjoint:
\beq
D^s_{\mu \nu}(p\cdot \sigma)
= \sum_{\alpha}D^s_{\mu\alpha}(\Lambda_c(p))D^s_{\alpha\nu}(\Lambda_c(p))^{\dagger} .
\label{d.17}
\eeq

The treatment of relativity follows from the relation between
the four dimensional Euclidean group and the associated complex
subgroup of the Lorentz group discussed in section 2.  Consider the two matrices
\beq
p \cdot \sigma :=
\left (
\begin{array}{cc}
p^0+p^2 & p^1 -i p^2\\
p^1+i p^2 & p^0 - p^3\\
\end{array}  
\right)  
\qquad 
p_e \cdot \sigma_e :=
\left (
\begin{array}{cc}
i p_e^0+p_e^2 & p_e^1 -i p_e^2\\
p_e^1+i p_e^2 & i p_e^0 - p_e^3\\
\end{array}  
\right).  
\label{d.18}
\eeq
The $SL(2,\mathbb{C})\times SL(2,\mathbb{C})$ transformation properties of
these matrices (denoted by $P$) are
\beq
P\to P' = A P B^t .
\label{d.19}
\eeq
The associated complex $4 \times 4 $ Lorentz and four-dimensional orthogonal
transformation
matrices are
\beq
\Lambda (A,B)^{\mu}{}_{\nu} = {1 \over 2} \mbox{Tr}(\sigma_{\mu}A\sigma_{\nu}B^t)
\qquad
{\cal O} (A,B)^{\mu}{}_{\nu} = {1 \over 2} \mbox{Tr}(\sigma_{e\mu}^{\dagger}A\sigma_{e\nu}B^t) .
\eeq
For ordinary rotations
$A=B^*=e^{i{\lambda \over 2} \hat{\mathbf{n}}}$.
For rotations about the $\hat{\mathbf{z}}$ axis 
\beq
{\cal O}(A,A^*)( \lambda) = 
\left (
\begin{array}{cccc}
1 & 0 & 0 & 0 \\
0 & \cos(\lambda)&\sin (\lambda)&0\\
0 & -\sin(\lambda)&\cos (\lambda)&0\\
0 & 0 & 0& 1\\
\end{array}  
\right )
\label{d.20}
\eeq
and
\beq
\Theta {\cal O}(A,A^*)( \lambda) \Theta = {\cal O}(A,A^*)( \lambda).
\label{d.21}
\eeq
For real rotations in Euclidean space-time planes $A=B^t=e^{i{\lambda \over 2} \hat{\mathbf{n}}\cdot \pmb{\sigma}}$.
For the case of the $x^0_e-\hat{\mathbf{z}}$ plane
\beq
{\cal O}(A,A^t)( \lambda) x= 
\left (
\begin{array}{cccc}
\cos(\lambda) & 0 & 0 & \sin (\lambda) \\
0 & 1 & 0 &0\\
0 & 0 & 1 &0\\
-\sin (\lambda) & 0 & 0& \cos(\lambda)\\
\end{array}  
\right )
\label{d.22}
\eeq
\and
\beq
\Theta {\cal O}^t(A,A^t)( \lambda) \Theta = {\cal O}(A,A^t)( \lambda).
\label{d.23}
\eeq
While ordinary 3-dimensional rotations are the same for
$p\cdot \sigma$ or $p_e \cdot \sigma_e$,  real rotations in
Euclidean space time planes are interpreted as rotationless Lorentz boosts 
with imaginary rapidity.        

These identifications imply the following algebraic relations between the
infinitesimal generators of the four dimensional orthogonal group and  
the Lorentz group:
 \beq
\mathbf{P}_m = \mathbf{P}_e \qquad J^{ij}_{m} = J^{ij}_{e}
\label{d.24}
\eeq
\beq
H_m = i H_e \qquad  K^{i}_{m} = -i J^{0i}_{e}
\label{d.25}
\eeq
Because of the factor of $i$, if the Euclidean generators are
self-adjoint operators on a representation of the Hilbert space, the
constructed Poincar\'e generators cannot be self-adjoint on that
representation of the Hilbert space.

In the spinless case ($s=0$) the identifications (\ref{d.20}-\ref{d.23})
result in the following expressions for the infinitesimal generators
of the Poincar\'e group on the Hilbert space with the time reflection:
\beq
H\Psi (x_e)= {\partial \over \partial x^0_e}  \Psi (x_e)
\qquad
\mathbf{P}\Psi (x_e)= -i {\partial \over \partial \mathbf{x}_e }  \Psi (x_e)
\label{d.26}
\eeq
\beq
\mathbf{J}\Psi (x_e)= -i \mathbf{x}
\times \pmb{\nabla}_x  \Psi (x_e)
\qquad
K^j \Psi (x_e)=( x^j {\partial \over \partial x^0_e}-
x^0_e {\partial \over \partial x^j}) \Psi (x_e).
\label{d.27}
\eeq
It is straightforward to demonstrate that these operators satisfy
the Poincar\'e
commutations relations (\ref{c.16}-\ref{c.18}).  For example
\beq
[K^i,H] = [ 
x^i {\partial \over \partial x_e^0} -
x^0_e {\partial \over \partial x^i},
{\partial \over \partial x^0_e}]=
i (-i {\partial \over \partial x^i}) =
i P^i
\label{d.28}
\eeq
which agrees with (\ref{c.18}). The other commutators can be checked similarly.

The Euclidean time reversal of the final state makes both the
Hamiltonian $H$ and the boost generators $\mathbf{K}$ formally
Hermitian with respect to the scalar product (\ref{d.7}).  The
non-trivial observation is that even an infinitesimal rotation in a Euclidean
space time plane can map functions with positive Euclidean time support
to functions that violate the support condition.  This maps
Hilbert space vectors out of the Hilbert space.
The resolution of this problem will be discussed in section 6.

To show the hermiticity of the rotationless boost generators
(\ref{d.27})
note that rotational invariance of the Euclidean Green's function in
Euclidean space-time planes means that the Euclidean rotation
generators commute with the Euclidean Green's function:
\beq
(-i x^i {\partial \over \partial x_e^0} +i
x^0_e {\partial \over \partial x^i})
S^0_e(x-y) =
S^0_e(x-y)
(-i y^i {\partial \over \partial y_e^0} +i
y^0_e {\partial \over \partial y^i}).
\label{d.29}
\eeq
Multiplying both sides by $i$ gives
\beq
( x^i {\partial \over \partial x_e^0} -
x^0_e {\partial \over \partial x^i})
S^0_e(x-y) =
S^0_e(x-y)
(y^i {\partial \over \partial y_e^0} -
y^0_e {\partial \over \partial y^i}).
\label{d.30}
\eeq
Next consider the inner product
\[
\langle f \vert K^i \vert g \rangle =
\]
\beq
\int d^4x d^4y f^* (\mathbf{x},-x_e^0 ) S^0_e(x-y)
(y^i {\partial \over \partial y_e^0} -
y^0_e {\partial \over \partial y^i}) g(\mathbf{y},y^0_e).
\label{d.31}
\eeq
Using (\ref{d.30}) in (\ref{d.31}) gives  
\beq
= \int d^4x d^4y f^* (\mathbf{x},-x_e^0 )
(x^i {\partial \over \partial x_e^0} -
x^0_e {\partial \over \partial x^i})
S^0_e(x-y)
g(\mathbf{y},y^0_e).
\label{d.32}
\eeq
Integrating by parts again gives
\beq
= -\int d^4x d^4y
(x^i {\partial \over \partial x_e^0} +
x^0_e {\partial \over \partial x^i})
(\theta f)^* (\mathbf{x},x_e^0 )
S^0_e(x-y)
g(\mathbf{y},y^0_e).
\label{d.33}
\eeq
Finally factoring the time reversal out of $f$ gives
\beq
-(x^i {\partial \over \partial x_e^0} +
x^0_e {\partial \over \partial x^i}) \theta 
f^* (\mathbf{x},x_e^0)=
\theta
\left ((x^i {\partial \over \partial x_e^0} -
x^0_e {\partial \over \partial x^i}) 
f^* (\mathbf{x},x_e^0) \right )
\label{d.34}
\eeq
which when used in (\ref{d.33}) gives
\[
\langle f \vert K^i \vert g \rangle =
\int d^4x d^4y f^* (\mathbf{x},-x_e^0 ) S^0_e(x-y)
(y^i {\partial \over \partial y_e^0} -
y^0_e {\partial \over \partial y^i}) g(\mathbf{y},y^0_e) =
\]
\beq
\int d^4x d^4y
\theta ((x^i {\partial \over \partial x_e^0} -
x^0_e {\partial \over \partial x^i}) f (\mathbf{x},x_e^0 ))^*
S^0_e(x-y)
g(\mathbf{y},y^0_e) =
\langle K^i f \vert  g \rangle. 
\label{d.35}
\eeq
This shows that $K^i$ is a Hermitian operator on this representation
of the Hilbert space.

The other non-trivial operator is the Hamiltonian (\ref{d.26}).
In this case
\[
\langle f \vert H \vert g \rangle =
\int d^4x d^4y f^* (\mathbf{x},-x_e^0 ) S^0_e(x-y)
{\partial \over \partial y_e^0}
g(\mathbf{y},y_e^0) =
\]
\[
-\int d^4x d^4y f^* (\mathbf{x},-x_e^0 )
{\partial \over \partial y_e^0}
S^0_e(x-y)
g(\mathbf{y},y^0_e) =
\int d^4x d^4y f^* (\mathbf{x},-x_e^0 )
{\partial \over \partial x_e^0}
S^0_e(x-y)
g(\mathbf{y},y^0_e) =
\]
\beq
-\int d^4x d^4y
{\partial \over \partial x_e^0}
f^* (\mathbf{x},-x_e^0 )
S^0_e(x-y)
g(\mathbf{y},y^0_e) =
\int d^4x d^4y
{\partial f^* \over \partial x^0}
(\mathbf{x},-x_e^0 ) 
S^0_e(x-y)
g(\mathbf{y},y^0_e) =
\langle Hf \vert g \rangle .
\label{d.36}
\eeq
The Euclidean time reversal does not change the linear or angular
momentum operators.  These methods can be used to demonstrate that all
of the $s=0$ generators (\ref{d.26}-\ref{d.27}) are Hermitian in this
representation of the Hilbert space and satisfy the Poincar\'e Lie
algebra.

\section{Spin}

In this section explicit formulas for generators for particles with
arbitrary spin are derived, generalizing the method used in the
previous section for scalar particles.  

In the original Poincar\'e covariant theory the spin is associated
with the observable that is the $\hat{\mathbf{z}}$-component of the
spin that would be
measured in the particle's rest frame if it was transformed to the
rest frame with a rotationless Lorentz transformation.  The spin in
the covariant wave function is related to this spin by multiplying
by one of the $SL(2,\mathbb{C})$ matrices, $D^s_{\mu \nu}(\Lambda_c(p)^{-1})$ or
$D^s_{\mu \nu}(\Lambda_c(p)^{\dagger})$.  These transformations lead to distinct
right or left handed spinors.  In discussing spin it is important to
understand that the Poincar\'e covariant spinors and the Lorentz
covariant spinors are related, but different.  Representations of the
Poincar\'e generators for each type of covariant spin must be
considered separately. In addition, for each type of covariant spinor
there are invariant linear functionals that define dual spinors.  The
dual spinors are spinor analogs of covariant and contravariant
vectors.  In conventional treatments
\cite{Wightman:1980}\cite{Wightman} \cite{berestetskii} the
right-handed spinors are denoted by $\xi^a$, left handed spinors are denoted
by $\xi^{\dot{a}}$ and their duals are denoted by $\xi_a$ and
$\xi_{\dot{a}}$ respectively.

The first step is to consider the $SL(2,\mathbb{C})$ transformation
properties of the Euclidean kernels for right and left handed
covariant spinors and their duals.

Euclidean four vectors can be represented by any of the four matrices:
\beq
p_e \cdot \sigma_e = p_e^\mu\sigma_{e\mu}
\qquad
p_e \cdot ( \sigma_2 \sigma_e \sigma_2 )  =  p_e^\mu\sigma_2 \sigma_{e\mu} \sigma_2
\qquad
p_e \cdot \sigma^t_e  =  p_e^\mu\sigma^t_{e\mu}
\qquad
p_e \cdot ( \sigma_2 \sigma_e^t \sigma_2 )
=  p_e^\mu\sigma_2 \sigma^t_{e\mu} \sigma_2 .
\label{s.1}
\eeq
The determinant of each of these matrices is (-) the square of the
Euclidean length of $p_e$, which is preserved under linear 
transformations of the form
\beq
P' = APB^t
\label{s.2}
\eeq
where $P$ represents any of the matrices in (\ref{s.1}), and $A,B \in
SL(2,\mathbb{C})$.  Real four-dimensional orthogonal transformations
are obtained by restricting $A$ and $B$ to be elements of $SU(2)$.

The $4 \times 4$ orthogonal matrix $\mathbb{O}(A,B)^{\mu}{}_{\nu}$
is related to the pair $(A,B)$ by
\beq
\mathbb{O}(A,B)^{\mu}{}_{\nu}:=\frac{1}{2}Tr(\sigma_{e\mu}^\dagger A\sigma_{e\nu}B^t).
\label{s.3}
\eeq
It follows that
\beq
A p_e^{\mu}\sigma_{e\mu} B^t= \sigma_{e\mu}\mathbb{O}(A,B)^{\mu}{}_{\nu}p_e^{\nu}
 = \sigma_{e\mu} (\mathbb{O}(A,B)p_e)^{\mu}.
\label{s.4}
\eeq
Multiplying
(\ref{s.4}) by $\sigma_2$ on both sides gives
\beq
A^*(p_e \cdot \sigma_2\sigma_e\sigma_2)B^\dagger=(\mathbb{O}(A,B)p)_e\cdot\sigma_2\sigma_e\sigma_2 .
\label{s.5}
\eeq
Taking transposes of the $2 \times 2$ matrices (\ref{s.4}) and (\ref{s.5})
give 
\beq
B(p_e\sigma_e^t)A^t= \sigma_e^{t}\cdot(\mathbb{O}(A,B)p_e) 
\label{s.6}
\eeq
and
\beq
B^*(p_e \cdot \sigma_2\sigma_e^{t}\sigma_2)A^\dagger=
\sigma_2\sigma_e^{t}\sigma_2 \cdot
(\mathbb{O}(A,B)p_e)
\label{s.7}
\eeq
where $\sigma_2 A \sigma_2 = A^*$ for $A\in SU(2)$ was used in
(\ref{s.5}-\ref{s.7}).  In all four of these expressions
$A$, $B$ and the orthogonal matrix
$\mathbb{O}(A,B)$ are unchanged.  All four of the
matrices (\ref{s.1}) become positive when $p_e$ is replaced by the
on-shell Minkowski four momentum,
$p^{\mu}_m = (\sqrt{\mathbf{p}^2+m^2},\mathbf{p})$
and $\sigma_e^{\mu}$ is replaced by $\sigma^{\mu}$. 

These matrices appear in the Euclidean covariant
kernels for the right and left-handed representations
and their duals.  The spin $s$ Euclidean covariant kernels
for each type of covariant spinor are:
\beq
S_e^s (x_e;\mu,\nu)={2 \over (2\pi)^4} \int {D^s_{\mu\nu} [p_e \cdot \sigma_{e}]\over p_e^2 +m^2}
e^{i p_e \cdot x_e}d^4p_e
\label{s.8}
\eeq
\beq
S_{ed}^s (x_e;\mu,\nu) ={2 \over (2\pi)^4} \int
{D^s_{\mu\nu}  [p_e \cdot (\sigma_2 \sigma_{e} \sigma_2)]\over p_e^2 +m^2}
e^{i p_e \cdot x_e}d^4p_e
\label{s.9}
\eeq
\beq
S_{e*}^s (x_e;\mu,\nu) ={2 \over (2\pi)^4} \int
{D^s_{\mu\nu}  [p_e \cdot \sigma^t_{e}]\over p_e^2 +m^2}
e^{i p_e \cdot x_e}d^4p_e
\label{s.10}
\eeq
\beq
S_{ed*}^s (x_e;\mu,\nu) ={2 \over (2\pi)^4} \int
{D^s_{\mu\nu}  [p_e \cdot (\sigma_2 \sigma^t_{e} \sigma_2)]\over p_e^2 +m^2}
e^{i p_e \cdot x_e}d^4p_e .
\label{s.11}
\eeq
The physical Hilbert space
inner product associated with each of these kernels is
\beq
\langle\psi_{e} \vert \phi_{e} \rangle=
\int \sum_{\mu \nu}\psi_e^*(\theta x,\mu)
S_e^s (x_e-y_e;\mu,\nu) \phi_e(y,\nu)d^4x d^4y
\label{s.12}
\eeq
\beq
\langle\psi_{ed} \vert \phi_{ed} \rangle=
\int \sum_{\mu \nu} \psi^*_{ed}(\theta x,\mu)
S_{ed}^s (x_e-y_e;\mu,\nu) \phi_{ed}(y,\nu)d^4x d^4y
\label{s.13}
\eeq
\beq
\langle\psi_{e*} \vert \phi_{e*} \rangle=
\int \sum_{\mu \nu} \psi^*_{e*}(\theta x,\mu)
S_{e*}^s (x_e-y_e;\mu,\nu) \phi_{e*}(y,\nu)d^4x d^4y
\label{s.14}
\eeq
\beq
\langle\psi_{ed*} \vert \phi_{ed*} \rangle=
\int \sum_{\mu \nu} \psi^*_{ed*}(\theta x,\mu)
S_{ed*}^s (x_e-y_e;\mu,\nu) \phi_{ed*}(y,\nu)d^4x d^4y .
\label{s.15}
\eeq
For wave functions with positive Euclidean time support, the $p^0_e$
integral can be evaluated by the residue theorem, closing the contour
in the lower half plane.  This replaces $p^0_e$ by
$-i\omega_m(\mathbf{p})$.  The kernels become the two-point
Minkowski Wightman functions \cite{Wightman:1980} for
mass $m$ spin $s$ irreducible
representations of the Lorentz group.  Equations (\ref{s.13}) and
(\ref{s.15}) are dual representations of the right-handed kernel,
while (\ref{s.12}) and (\ref{s.14}) are dual representations of the
left-handed kernel.  $\sigma_2$ behaves like a metric tensor for the
Lorentz covariant spinors, relating the representations
(\ref{s.12}) and (\ref{s.13}) or (\ref{s.14}) and (\ref{s.15}).
Contraction of the two types of right
or left handed spinors are Lorentz invariant.  The results of performing
the
$p^0_e$ integral are
\beq
\langle \psi_{e} \vert \phi_{e}\rangle= 
\int \sum_{\mu \nu} f^*_m(\mathbf{p},\mu)
{d\mathbf{p}
D_{\mu \nu}^s[p_m\cdot
\sigma] \over \omega_m (\mathbf{p})}
g_m(\mathbf{p},\nu)
\label{s.16}
\eeq
\beq
\langle \psi_{ed} \vert \phi_{ed} \rangle=
\int \sum_{\mu \nu} f^*_m(\mathbf{p},\mu)
{d\mathbf{p}
D_{\mu \nu}^s[p_m\cdot \sigma_2 \sigma
\sigma_2] \over \omega_m (\mathbf{p})}
g_m(\mathbf{p},\nu)
\label{s.17}
\eeq
\beq
\langle \psi_{e*}\vert \phi_{e*}\rangle=
\int \sum_{\mu \nu}
f^*_m(\mathbf{p},\mu)
{d\mathbf{p}
D_{\mu \nu}^s[p_m\cdot
\sigma^*] \over \omega_m (\mathbf{p})}
g_m(\mathbf{p},\nu)
\label{s.18}
\eeq
\beq
\langle \psi_{ed*} \vert \phi_{ed*}\rangle=
\int \sum_{\mu \nu} f^*_m(\mathbf{p},\mu)
{d\mathbf{p}
D_{\mu \nu}^s[p_m\cdot \sigma_2
\sigma^* \sigma_2 ]  
\over \omega_m (\mathbf{p})}
g_m(\mathbf{p},\nu)
\label{s.19}
\eeq
where
\beq
f^*_m(\mathbf{p},\mu):= 
\int {d^4x \over (2 \pi)^{3/2}}
\psi^*(x,\mu)e^{i \mathbf{p}\cdot \mathbf{x} - \omega_m (\mathbf{p})x^0}
\label{s.20}
\eeq
\beq
g_m(\mathbf{p},\nu):=
{d^4y \over (2 \pi)^{3/2}}
\psi(y,\nu)e^{-i \mathbf{p}\cdot \mathbf{y} - \omega_m (\mathbf{p})y^0}
\label{s.21}
\eeq
for each type of spinor wave function. 

Each of the spin matrices, 
$D_{\mu \nu}^s[p_m\cdot \sigma]$,
$D_{\mu \nu}^s[p_m\cdot \sigma_2 \sigma \sigma_2]$,
$D_{\mu \nu}^s[p_m\cdot \sigma^*]$ and 
$D_{\mu \nu}^s[p_m\cdot \sigma_2 \sigma^* \sigma_2 ]$ are   
positive Hermitian matrices,
so the Euclidean Green's functions (\ref{s.8}-\ref{s.11})
are all reflection positive.

The first step to find the spinor parts of the Poincar\'e generators
in the Euclidean representation is 
to use the identities (\ref{s.4}-\ref{s.7}) which lead to 
\[
\int \sum_{\mu \nu} \psi_e^*(\theta x,\mu) \frac{e^{ip\cdot(x-y)}}{p^2+m^2}D_{\mu \nu}^s[\mathbb{O}p\cdot \sigma_e]\phi_e(y,\nu)d^4x d^4y d^4p=
\]
\beq
\int \sum_{\mu \nu} \psi_e^*(\theta x,\mu) \frac{e^{ip\cdot(x-y)}}{p^2+m^2}D_{\mu \nu}^s[p\cdot A\sigma_e B^t]\phi_e(y,\nu)d^4x d^4y d^4p
\label{s.22}
\eeq
\[
\int \sum_{\mu \nu} \psi_{ed}^*(\theta x,\mu) \frac{e^{ip\cdot(x-y)}}{p^2+m^2}
D_{\mu \nu}^s[\mathbb{O}p\cdot \sigma_2 \sigma_e \sigma_2]
\phi_{ed}(y,\nu)d^4x d^4y d^4p=
\]
\beq
\int \sum_{\mu \nu} \psi_{ed}^*(\theta x,\mu) \frac{e^{ip\cdot(x-y)}}{p^2+m^2}D_{\mu \nu}^s[p\cdot A^*\sigma_2 \sigma_e \sigma_2B^\dagger]\phi_{ed}(y,\nu)d^4x d^4y d^4p
\label{s.23}
\eeq

\[
\int \sum_{\mu \nu} \psi^*_{e*}(\theta x,\mu) \frac{e^{ip\cdot(x-y)}}{p^2+m^2}D_{\mu \nu}^s[\mathbb{O}p\cdot \sigma_e^t]\phi_{e*}(y,\nu)d^4x d^4y d^4p=
\]
\beq
\int \sum_{\mu \nu} \psi^*_{e*}(\theta x,\mu) \frac{e^{ip\cdot(x-y)}}{p^2+m^2}D_{\mu \nu}^s[p\cdot B\sigma_e^tA^t]\phi_{e*}(y,\nu)d^4x d^4y d^4p
\label{s.24}
\eeq
 
\[
\int \sum_{\mu \nu} \psi^*_{ed*}(\theta x,\mu) \frac{e^{ip\cdot(x-y)}}{p^2+m^2}D_{\mu \nu}^s[\mathbb{O}p\cdot \sigma_2 \sigma_e^t \sigma_2 ]\phi_{ed*}(y,\nu)d^4x d^4y d^4p=
\]
\beq
\int \sum_{\mu \nu} \psi_{ed*}^*(\theta x,\mu) \frac{e^{ip\cdot(x-y)}}{p^2+m^2}D_{\mu \nu}^s[p\cdot B^*\sigma_2 \sigma_e^t \sigma_2A^\dagger ]\phi_{ed*}(y,\nu)d^4x d^4y d^4p .
\label{s.25}
\eeq

The next step is to move the transformations from the kernels to the
wave functions.  The Euclidean invariance of the measures and scalar
products, the group representation properties of the Wigner functions,
and re-definitions of the wave functions can be used to show that
(\ref{s.22}-\ref{s.25}) are equivalent to
\[
\int \sum   (D^s_{\mu\alpha}[A^\dagger]^{-1}{\psi}_e(\theta \mathbb{O}^t \theta  x,\alpha))^* \frac{e^{ip\cdot(\theta x-y)}}{p^2+m^2}D_{\mu \nu}^s[p\cdot \sigma_e]
{\phi}_e(y,\nu)d^4x d^4y d^4p
\]
\beq
=\int \sum  {\psi}_e^*( x,\mu) \frac{e^{ip\cdot(\theta x-y)}}{p^2+m^2}D_{\mu \alpha}^j[p\cdot \sigma_e]D^s_{\alpha \nu}[B^t]{\phi}_e(\mathbb{O}y,\nu)d^4x d^4y d^4p
\label{s.26}
\eeq

\[
\int \sum  ( D^s_{\mu \alpha}[A^t]^{-1}{\psi}_{ed}(\theta \mathbb{O}^t \theta   x,\alpha))^* \frac{e^{ip\cdot(\theta x-y)}}{p^2+m^2}D_{\mu \nu}^s[p\cdot \sigma_2 \sigma_e \sigma_2]{\phi}_{ed}(y,\nu)d^4x d^4y d^4p
\]
\beq
=\int \sum  {\psi}^*_{ed}( x,\mu) \frac{e^{ip\cdot(\theta x-y)}}{p^2+m^2}D_{\mu \alpha}^j[p\cdot \sigma_2 \sigma_e \sigma_2]
D^s_{\alpha \nu}[B^\dagger]{\phi}_{ed}(\mathbb{O}y,\nu)d^4x d^4y d^4p
\label{s.27}
\eeq

\[
\int \sum  (D^s_{\mu \alpha}[B^\dagger]^{-1} {\psi}_{e*}(\theta  \mathbb{O}^t \theta x,\alpha))^* \frac{e^{ip\cdot(\theta x-y)}}{p^2+m^2}D_{\mu \nu}^s[p\cdot \sigma_e^t]{\phi}_{e*}(y,\nu)d^4x d^4y d^4p
\]
\beq
=\int \sum  {\psi}_{e*}^*( x,\mu) \frac{e^{ip\cdot(\theta x-y)}}{p^2+m^2}D_{\mu \alpha}^j[p\cdot \sigma_e^t]D^s_{\alpha \nu}[A^t]\Tilde{\phi}_{e*}(\mathbb{O}y,\nu)d^4x d^4y d^4p
\label{s.28}
\eeq

\[
\int \sum  (D^s_{\mu \alpha}[B^t]^{-1} {\psi}_{ed*}(\theta  \mathbb{O}^t\theta  x,\alpha))^* \frac{e^{ip\cdot(\theta x-y)}}{p^2+m^2}D_{\mu \nu}^s[p\cdot \sigma_2 \sigma_e^t \sigma_2 ]{\phi}_{ed*}(y,\nu)d^4x d^4y d^4p
\]
\beq
= \int \sum  {\psi}_{ed*}^*( x,\mu) \frac{e^{ip\cdot(\theta
    x-y)}}{p^2+m^2}D_{\mu \alpha}^s[p\cdot \sigma_2 \sigma_e^t \sigma_2]
D^s_{\alpha \nu}[A^\dagger]{\phi}_{ed*}(\mathbb{O}y,\nu)d^4x d^4y d^4p .
\label{s.29}
\eeq
To derive expressions for the generators for each type of spinor,
check the hermiticity and verify the commutation relations
the first step is to replace $A$ and $B$ with
the pairs of $SU(2)$ matrices representing 
one-parameter groups for both ordinary rotations about a fixed
axis and rotations in a Euclidean space time plane.

For ordinary rotations about the $\hat{\mathbf{n}}$ axis,
the one-parameter group is 
\beq
A(\lambda ) = B^*(\lambda) = e^{i {\lambda \over 2}\hat{\mathbf{n}}
\cdot \pmb{\sigma}}
\label{s.30}
\eeq
and
$(\theta  \mathbb{O}^t(\lambda)\theta) = \mathbb{O}^t(\lambda)$,
while for rotations in Euclidean $\hat{\mathbf{n}}$-$x^0$  space-time planes
the one-parameter group is
\beq
A(\lambda ) = B^t(\lambda) = e^{i {\lambda \over 2}\hat{\mathbf{n}}
\cdot \pmb{\sigma}}
\label{s.31}
\eeq
and
$(\theta  \mathbb{O}^t(\lambda)\theta) = \mathbb{O}(\lambda)$.
The $4 \times 4$ orthogonal transformations, $\mathbb{O}(\lambda)$
associated with each
type of transformation are shown explicitly for
rotations about the $\hat{\mathbf{z}}$ axis and for
rotations in the $\hat{\mathbf{z}}$-$x^0$
plane:
For rotations about the $\hat{\mathbf{z}}$ axis 
\beq
\mathbb{O}(A,A^*)( \lambda) = 
\left (
\begin{array}{cccc}
1 & 0 & 0 & 0 \\
0 & \cos(\lambda)&\sin (\lambda)&0\\
0 & -\sin(\lambda)&\cos (\lambda)&0\\
0 & 0 & 0& 1\\
\end{array}  
\right )
\label{s.32}
\eeq
and
\beq
\Theta \mathbb{O}(A,A^*)( \lambda) \Theta = \mathbb{O}(A,A^*)( \lambda).
\label{s.33}
\eeq
For rotations in the $\hat{\mathbf{z}}$-$x^0$ plane
\beq
\mathbb{O}(A,A^t)( \lambda)= 
\left (
\begin{array}{cccc}
\cos(\lambda) & 0 & 0 & \sin (\lambda) \\
0 & 1 & 0 &0\\
0 & 0 & 1 &0\\
-\sin (\lambda) & 0 & 0& \cos(\lambda)\\
\end{array}  
\right )
\label{s.34}
\eeq
and
\beq
\theta \mathbb{O}^t(A,A^t)( \lambda) \theta = \mathbb{O}(A,A^t)( \lambda)
\label{s.35}
\eeq

For the case of ordinary rotations
$A=B^*$ and equations (\ref{s.26}-\ref{s.29}) become
\[
\int \sum  (D^s_{\mu\alpha}[A]{\psi}_e ( \mathbb{O}^t(\lambda)   x,\alpha) )^*  \frac{e^{ip\cdot(\theta x-y)}}{p^2+m^2}D_{\mu \nu}^s[p\cdot \sigma_e]{\phi}_e(y,\nu)d^4x d^4y d^4p
\]
\beq
=\int \sum  {\psi}_e^*( x,\mu) \frac{e^{ip\cdot(\theta x-y)}}{p^2+m^2}D_{\mu \alpha}^j[p\cdot \sigma_e]D^s_{\alpha \nu}[A^{\dagger}]{\phi}_e(\mathbb{O}(\lambda)y,\nu)d^4x d^4y d^4p
\label{s.36}
\eeq

\[
  \int \sum ( D^s_{\mu \alpha}[A^*]{\psi}_{ed}( \mathbb{O}^t(\lambda)   x,\alpha))^*
  \frac{e^{ip\cdot(\theta x-y)}}{p^2+m^2}D_{\mu \nu}^s[p\cdot \sigma_2 \sigma_e \sigma_2]{\phi}_{ed}(y,\nu)d^4x d^4y d^4p
\]
\beq
=\int \sum  {\psi}_{ed}^*( x,\mu) \frac{e^{ip\cdot(\theta x-y)}}{p^2+m^2}D_{\mu \alpha}^j[p\cdot \sigma_2 \sigma_e \sigma_2]
D^s_{\alpha \nu}[A^t] {\phi}_{ed}(\mathbb{O}(\lambda)y,\nu)d^4x d^4y d^4p
\label{s.37}
\eeq

\[
\int \sum  (D^s_{\mu \alpha}[A^*] {\psi}_{e*}( \mathbb{O}^t(\lambda)   x,\alpha))^*
\frac{e^{ip\cdot(\theta x-y)}}{p^2+m^2}D_{\mu \nu}^s[p\cdot \sigma_e^t]{\phi}_{e*}(y,\nu)d^4x d^4y d^4p
\]
\beq
=\int \sum  {\psi}_{e*}^*( x,\mu) \frac{e^{ip\cdot(\theta x-y)}}{p^2+m^2}D_{\mu \alpha}^j[p\cdot \sigma_e^t]D^s_{\alpha \nu}[A^t]{\phi}_{e*}(\mathbb{O}(\lambda)y,\nu)d^4x d^4y d^4p
\label{s.38}
\eeq

\[
\int \sum (D^s_{\mu \alpha}[A] {\psi}_{ed*}( \mathbb{O}^t(\lambda)   x,\alpha))^*
\frac{e^{ip\cdot(\theta x-y)}}{p^2+m^2}D_{\mu \nu}^s[p\cdot \sigma_2
\sigma_e^t \sigma_2 ]{\phi}_{ed*}(y,\nu)d^4x d^4y d^4p
\]
\beq
= \int \sum {\psi}_{ed*}^*( x,\mu) \frac{e^{ip\cdot(\theta
    x-y)}}{p^2+m^2}D_{\mu \alpha}^s[p\cdot \sigma_2 \sigma_e^t \sigma_2
]D^s_{\alpha \nu}[A^\dagger]{\phi}_{ed*}(\mathbb{O}(\lambda)y,\nu)d^4x d^4y d^4p
\label{s.39}
\eeq

For the case of rotations in Euclidean space-time planes
for $A=B^t$ equations (\ref{s.26}-\ref{s.29}) become
\[
\int \sum  (D^s_{\mu\alpha}[A]{\psi}_e(\mathbb{O}(\lambda)  x,\alpha))^* \frac{e^{ip\cdot(\theta x-y)}}{p^2+m^2}D_{\mu \nu}^s[p\cdot \sigma_e]{\phi_e}(y,\nu)d^4x d^4y d^4p
\]
\beq
=\int \sum  {\psi_e}^*( x,\mu) \frac{e^{ip\cdot(\theta x-y)}}{p^2+m^2}D_{\mu \alpha}^j[p\cdot \sigma_e]D^s_{\alpha \nu}[A]
{\phi_e}(\mathbb{O}(\lambda)y,\nu)d^4x d^4y d^4p
\label{s.40}
\eeq

\[
\int \sum ( D^s_{\mu \alpha}[A^*]{\psi}_{ed}(\mathbb{O}(\lambda)  x,\alpha))^* \frac{e^{ip\cdot(\theta x-y)}}{p^2+m^2}D_{\mu \nu}^s[p\cdot \sigma_2 \sigma_e \sigma_2]{\phi}_{ed}(y,\nu)d^4x d^4y d^4p
\]
\beq
=\int \sum {\psi}_{ed}^*( x,\mu) \frac{e^{ip\cdot(\theta x-y)}}{p^2+m^2}D_{\mu \alpha}^j[p\cdot \sigma_2 \sigma_e \sigma_2]
D^s_{\alpha \nu}[A^*]{\phi}_{ed}(\mathbb{O}y,\nu)d^4x d^4y d^4p
\label{s.41}
\eeq

\[
\int \sum (D^s_{\mu \alpha}[A^t] {\psi}_{e*}(\mathbb{O}(\lambda)  x,\alpha))^* \frac{e^{ip\cdot(\theta x-y)}}{p^2+m^2}D_{\mu \nu}^s[p\cdot \sigma_e^t]{\phi}_{e*}(y,\nu)d^4x d^4y d^4p
\]
\beq
=\int \sum {\psi}_{e*}^*( x,\mu) \frac{e^{ip\cdot(\theta x-y)}}{p^2+m^2}D_{\mu \alpha}^j[p\cdot \sigma_e^t]D^s_{\alpha \nu}[A^t]{\phi}_{e*}(\mathbb{O}(\lambda)y,\nu)d^4x d^4y d^4p
\label{s.42}
\eeq

\[
\int \sum (D^s_{\mu \alpha}[A^\dagger ] {\psi}_{ed*}(\mathbb{O}(\lambda)  x,\alpha))^*
\frac{e^{ip\cdot(\theta x-y)}}{p^2+m^2}D_{\mu \nu}^s[p\cdot \sigma_2 \sigma_e^t \sigma_2 ]{\phi}_{ed*}(y,\nu)d^4x d^4y d^4p
\]
\beq
= \int \sum  {\psi}_{ed*}^*( x,\mu) \frac{e^{ip\cdot(\theta
    x-y)}}{p^2+m^2}D_{\mu \alpha}^s[p\cdot \sigma_2 \sigma_e^t \sigma_2
]D^s_{\alpha \nu}[A^\dagger]{\phi}_{ed*}(\mathbb{O}(\lambda)y,\nu)d^4x d^4y d^4p.
\label{s.43}
\eeq

Note that the transformations above represent inverse Lorentz
transformations since
\beq
\langle x, \nu \vert U(\Lambda,0) \vert \psi \rangle =
\langle \psi \vert U^{\dagger}(\Lambda,0)\vert x,\nu \rangle^* = 
\langle \mathbb{O}^t x,\nu  \vert \psi \rangle^*=
\langle \psi \vert \mathbb{O}^t x, \nu \rangle .
\label{s.44}
\eeq
To construct generator of ordinary rotations differentiate the
right hand side of
(\ref{s.36}-\ref{s.39}) by $\lambda$, set $\lambda=0$, and multiply the result
by $i$.   To construct the generators of Euclidean space-time rotations
differentiate the right hand side of
(\ref{s.40}-\ref{s.43}) by $\lambda$, set $\lambda=0$, and multiply the
result by $i$
to get expressions for the generators.  To get expressions for the
Lorentz Boost generators multiply the Euclidean space-time rotation
generators by an additional factor of $-i$. 
The derivatives of the
Wigner functions can be computed using
\beq
{d \over d\lambda}D^s_{\mu \nu} [A(\lambda)]_{\vert_{\lambda=0}} =
{d \over d\lambda}\langle s, \mu \vert
e^{i \lambda \hat{\mathbf{n}}\cdot \mathbf{S}}
\vert s, \nu \rangle_{\vert_{\lambda=0}}=
i \langle s, \mu \vert \hat{\mathbf{n}}\cdot \mathbf{S} \vert s, \nu \rangle
\label{s.45}
\eeq
\beq
{d \over d\lambda}D^s_{\mu \nu} [A(\lambda)^{\dagger}]_{\vert_{\lambda=0}} =
{d \over d\lambda}\langle s, \mu \vert e^{-i \lambda \hat{\mathbf{n}}\cdot \mathbf{S}} \vert s, \nu \rangle_{\vert_{\lambda=0}} =
- i \langle s, \mu \vert \hat{\mathbf{n}}\cdot \mathbf{S} \vert s, \nu \rangle
\label{s.46}
\eeq
\beq
{d \over d\lambda}D^s_{\mu \nu} [A^*(\lambda)]_{\vert_{\lambda=0}} =
{d \over d\lambda}(D^s_{\mu \nu} [A(\lambda)])^*_{\vert_{\lambda=0}} =
-i \langle s, \mu \vert \hat{\mathbf{n}}\cdot \mathbf{S} \vert s, \nu \rangle^*=
-i \langle s, \nu \vert \hat{\mathbf{n}}\cdot \mathbf{S} \vert s, \mu \rangle
\label{s.47}
\eeq
\beq
{d \over d\lambda}D^s_{\mu \nu} [A^t(\lambda)]_{\vert_{\lambda=0}} =
{d \over d\lambda}(D^s_{\mu \nu} ([A(\lambda)])^*)^{-1}_{\vert_{\lambda=0}} =
i\langle s, \mu \vert \hat{\mathbf{n}}\cdot \mathbf{S} \vert s, \nu \rangle^*=
i\langle s, \nu \vert \hat{\mathbf{n}}\cdot \mathbf{S} \vert s, \mu \rangle .
\label{s.48}
\eeq
These can be evaluated using $S_z$ and angular momentum
raising and lowering operators.
The rotation generators for each type of spinor representation can be read off
of (\ref{s.36}-\ref{s.39}):
\beq
\langle x,s,\nu \vert  \mathbf{J} \vert \psi_e \rangle =
\sum_\nu \left (\delta_{\mu \nu} \mathbf{x} \times (-i {\partial \over \partial \mathbf{x}})
+ \langle s, \mu \vert \hat{\mathbf{n}}\cdot \mathbf{S} \vert s, \nu \rangle
\right ) \langle x,s,\nu \vert \psi_e \rangle 
\label{s.49}
\eeq
\beq
\langle x,s,\nu \vert  \mathbf{J} \vert \psi_{ed} \rangle =  
\sum_\nu \left ( \delta_{\mu \nu} \mathbf{x} \times (-i {\partial \over \partial \mathbf{x}})
- \langle s, \nu \vert \hat{\mathbf{n}}\cdot \mathbf{S} \vert s, \mu \rangle
\right ) \langle x,s,\nu \vert \psi_{ed} \rangle
\label{s.50}
\eeq
\beq
\langle x,s,\nu \vert  \mathbf{J} \vert \psi_{e*} \rangle =  
\sum_\nu \left ( \delta_{\mu \nu} \mathbf{x} \times (-i {\partial \over \partial \mathbf{x}})
- \langle s, \nu \vert \hat{\mathbf{n}}\cdot \mathbf{S} \vert s, \mu \rangle
\right ) \langle x,s,\nu \vert \psi_{e*} \rangle
\label{s.51}
\eeq
\beq
\langle x,s,\nu \vert  \mathbf{J} \vert \psi_{ed*} \rangle =  
\sum_\nu \left ( \delta_{\mu \nu} \mathbf{x} \times (-i {\partial \over \partial \mathbf{x}})
+ \langle s, \mu \vert \hat{\mathbf{n}}\cdot \mathbf{S} \vert s, \nu \rangle
\right ) \langle x,s,\nu \vert \psi_{ed*} \rangle
\label{s.52}
\eeq
The first and fourth term are representations of standard rotation generators
.  In the second and third terms the spin generator matrix elements are
transposed and multiplied by
with a (-) sign.  To show that these operator satisfy $SU(2)$
commutation relations, 
consider matrices satisfying $SU(2)$ commutation relations: 
\beq
[ M_i, M_j] = i \epsilon_{ijk} M_k . 
\label{s.53}
\eeq
The transposes satisfy
\beq
[ M^t_j, M^t_i] = i \epsilon_{ijk} M^t_k
\label{s.54}
\eeq
\beq
[ (-M^t_i),(- M^t_j)] = i \epsilon_{ijk} (-M^t_k)
\label{s.55}
\eeq
which shows that the negative transpose of these matrices also
satisfy $SU(2)$ commutation relations.  This shows that all of the
spin generator satisfy $SU(2)$ commutation relations.

Generators for rotations in
Euclidean space-time planes are constructed the same way from
\beq
\langle x,s,\nu \vert  J^{0\hat{n}} \vert \psi_{e} \rangle =   
\sum_\nu \left (i \delta_{\mu \nu} (\mathbf{x} {\partial \over \partial x^0} -
x^0 {\partial \over \partial \mathbf{x}}) 
- \langle s, \mu \vert \hat{\mathbf{n}}\cdot \mathbf{S} \vert s,\nu \rangle 
\right ) \langle x,s,\nu \vert \psi_e \rangle
\label{s.56}
\eeq
\beq
\langle x,s,\nu \vert  J^{0\hat{n}} \vert \psi_{ed} \rangle =
\sum_\nu \left (i \delta_{\mu \nu}(\mathbf{x} {\partial \over \partial x^0} -
x^0 {\partial \over \partial \mathbf{x}}) +
\langle s, \nu \vert \hat{\mathbf{n}}\cdot \mathbf{S} \vert s,\mu \rangle
\right ) \langle x,s,\nu \vert \psi_{ed} \rangle
\label{s.57}
\eeq
\beq
\langle x,s,\nu \vert  J^{0\hat{n}} \vert \psi_{e*} \rangle =  
\sum_\nu \left ( i \delta_{\mu \nu}(\mathbf{x} {\partial \over \partial x^0} -
x^0 {\partial \over \partial \mathbf{x}}) -
 \langle s, \nu \vert \hat{\mathbf{n}}\cdot \mathbf{S} \vert s,\mu \rangle 
\right ) \langle x,s,\nu \vert \psi_{e*} \rangle
\label{s.58}
\eeq
\beq
\langle x,s,\nu \vert  J^{0\hat{n}} \vert \psi_{ed*} \rangle =  
\sum_\nu \left ( i \delta_{\mu \nu}(\mathbf{x} {\partial \over \partial x^0} -
x^0 {\partial \over \partial \mathbf{x}}) 
+ \langle s, \mu \vert \hat{\mathbf{n}}\cdot \mathbf{S} \vert s,\nu \rangle 
\right ) \langle x,s,\nu \vert \psi_{ed*} \rangle
\label{s.59}
\eeq
In order to construct the boost generators it is necessary to multiply these
expression by an additional factor of (-i)
\beq
\langle x,s,\nu \vert  \mathbf{K}  \vert \psi_{e} \rangle =  
\sum_\nu \left ( \delta_{\mu \nu}(\mathbf{x} {\partial \over \partial x^0} -
x^0 {\partial \over \partial \mathbf{x}}) +
i \langle s, \mu \vert \hat{\mathbf{n}}\cdot \mathbf{S} \vert s,\nu \rangle 
\right ) \langle x,s,\nu \vert \psi_e \rangle
\label{s.60}
\eeq
\beq
\langle x,s,\nu \vert  \mathbf{K}  \vert \psi_{ed} \rangle =    
\sum_\nu \left ( \delta_{\mu \nu}(\mathbf{x} {\partial \over \partial x^0} -
x^0 {\partial \over \partial \mathbf{x}}) -i
\langle s, \nu \vert \hat{\mathbf{n}}\cdot \mathbf{S} \vert s,\mu \rangle
\right ) \langle x,s,\nu \vert \psi_{ed} \rangle
\label{s.61}
\eeq
\beq
\langle x,s,\nu \vert  \mathbf{K}  \vert \psi_{e*} \rangle =    
\sum_\nu \left ( \delta_{\mu \nu}(\mathbf{x} {\partial \over \partial x^0} -
x^0 {\partial \over \partial \mathbf{x}}) +i
 \langle s, \nu \vert \hat{\mathbf{n}}\cdot \mathbf{S} \vert s,\mu \rangle 
\right ) \langle x,s,\nu \vert \psi_{e*} \rangle
\label{s.62}
\eeq
\beq
\langle x,s,\nu \vert  \mathbf{K}  \vert \psi_{ed*} \rangle =    
\sum_\nu \left ( \delta_{\mu \nu} (\mathbf{x} {\partial \over \partial x^0} -
x^0 {\partial \over \partial \mathbf{x}}) 
-i \langle s, \mu \vert \hat{\mathbf{n}}\cdot \mathbf{S} \vert s,\nu \rangle 
\right ) \langle x,s,\nu \vert \psi_{ed*} \rangle
\label{s.63}
\eeq
The continuous part of these expressions agree with (\ref{d.26}-\ref{d.27})
for spinless operators.  The relevant commutators involving
the spin parts of the boost generators in each of the
four representations are
\beq
[K_i, K_j ] = [i S_i,i S_j]  = -i \epsilon_{ijk} S_k
\label{s.65}
\eeq
\beq
[K_i, K_j ] = [-i S^t_i,-i S^t_j] = - \epsilon_{ijk} (-S^t_k)
\label{s.65}
\eeq
\beq
[K_i, K_j ] = [i S^t_i,i S^t_j] = - \epsilon_{ijk} (-S^t_k)
\label{s.66}
\eeq
\beq
[K_i, K_j ] = [-i S_i,-i S_j]  = -i \epsilon_{ijk} S_k 
\label{s.67}
\eeq
\beq
[K_i, S_j ] = [i S_i, S_j]  =  \epsilon_{ijk}  iS_k = \epsilon_{ijk}  K_k
\label{s.68}
\eeq
\beq
[K_i, S_j ] = [-iS_i^t, -S^t_j]  =  \epsilon_{ijk}  -iS^t_k = \epsilon_{ijk}  K_k
\label{s.69}
\eeq
\beq
[K_i, S_j ] = [i S^t_i, -S^t_j]  =  \epsilon_{ijk}  iS^t_k = \epsilon_{ijk}  K_k
\label{s.70}
\eeq
\beq
[K_i, S_j ] = [-i S_i,S_j]  = -i \epsilon_{ijk} (-iS_k) = \epsilon_{ijk}  K_k
\label{s.71}
\eeq
where the spin generators in (\ref{s.65},\ref{s.66},\ref{s.69}) and
(\ref{s.70}) are (-) the transposes of the matrices satisfying $SU(2)$
commutation relations, which were shown in (\ref{s.53}-\ref{s.55}) to
satisfy $SU(2)$ commutation relations.  It follows that the
expressions (\ref{s.49}-\ref{s.52}) and (\ref{s.60}-\ref{s.63}) for the
Lorentz generators in each of the spinor representations satisfy the
Poincar\'e commutation relations.

The hermiticity of these generators follow from the expressions
(\ref{s.36}-\ref{s.39}) and (\ref{s.40}-\ref{s.43}).
Each of equations (\ref{s.36}-\ref{s.39}) has the form
\beq
\langle U^{\dagger}(\lambda)\psi \vert \phi \rangle = 
\langle \psi \vert U(\lambda) \vert \phi \rangle 
\label{s.72}
\eeq
so the rotation operators, which are generators of
unitary one-parameter groups \cite{riesz} are self-adjoint in the
Hilbert spaces with inner products (\ref{s.12}-\ref{s.15}).

For the boost generators hermiticity follows from
(\ref{s.40}-\ref{s.43}).  In this case all of
these equations have the form
\beq
\langle T(\lambda)\psi \vert \phi \rangle = 
\langle \psi \vert T(\lambda) \vert \phi \rangle  
\label{s.73}
\eeq
In these cases $T(\lambda)$ is Hermitian, but the generators are
constructed by multiplying the $\lambda$ derivative $1=(i)(-i)$
rather than $i$, resulting in Hermitian operators.

In these covariant representations the spin does not enter in
the Hamiltonian or the linear momentum operators.   These operators
all commute with the spin operators and commutators with these
operators follow from the scalar case. 

The main result of this section the expressions
(\ref{s.49}-\ref{s.52}) and (\ref{s.60}-\ref{s.63})
for the Poincar\'e generators.  The construction relates the
Euclidean spinors to the Lorentz covariant spinors.

\section{Self Adjointness} 

While the self-adjointness of the generators of ordinary rotations
follows from the unitarity of the one-parameter group of rotations on
the Hilbert spaces (\ref{s.12}-\ref{s.15}), this argument does not
apply to either the Hamiltonian or the boost generators.  In both
cases the operators were derived from the corresponding Euclidean
generators by multiplication by an imaginary constant.  The Euclidean
generators and corresponding Lorentz generators act on different
Hilbert space representations.  The problem is that the corresponding
finite Euclidean transformations can map functions with positive time
support to functions that violate this condition.

For the Hamiltonian this can be treated by only considering
translations in the positive Euclidean time direction.
These translations map functions with positive Euclidean time
support into functions with positive Euclidean time support.
Reflection positivity can be used to show that
translations in the positive Euclidean time direction define a
contractive Hermitian semigroup on the Hilbert space with the scalar products
(\ref{d.7}-\ref{d.10}).
The argument \cite{glimm} uses the Schwartz inequality on both
the physical and Euclidean Hilbert spaces.  One application of
the Schwartz inequality on the physical Hilbert space gives
\beq
\Vert \vert e^{-H x^0} \vert \phi \rangle \Vert
= 
\langle e^{-H x^0} \phi \vert e^{-H x^0} \vert \phi \rangle^{1/2}
=
\langle \phi \vert e^{-H 2x^0} \vert \phi \rangle^{1/2} \leq
\Vert \vert e^{-H 2x^0} \vert \phi \rangle \Vert^{1/2}
\Vert \vert \phi \rangle \Vert^{1/2} .
\label{sa.1}
\eeq
Repeating these steps $n$-times gives
\beq
\Vert \vert e^{-H x^0} \vert \phi \rangle \Vert \leq
\Vert \vert e^{-H 2^nx^0} \vert \phi \rangle \Vert^{1/2^n}
\Vert \vert \phi \rangle \Vert^{1-1/2^n} .
\label{sa.2}
\eeq
The quantity
\beq
\Vert \vert e^{-H 2^nx^0} \vert \phi \rangle \Vert \leq
\Vert \theta U_e(2^nx^0) \vert \psi \rangle \Vert_e <
\Vert \vert \psi \rangle \Vert_e <
\infty
\label{sa.3}
\eeq
is bounded by the Euclidean norm, $\Vert \cdot \Vert_e$, 
since $U_e(2^nx^0)$ is unitary and $\Vert \theta \Vert_e= 1$ on
that Hilbert space.
Since this is finite and independent of $n$, taking the limit as
$n \to \infty$ gives
\beq
\Vert \vert e^{-H x^0} \vert \phi \rangle \Vert \leq
\Vert \vert \phi \rangle \Vert .
\label{sa.4}
\eeq
It follows that positive Euclidean time translations define a
contractive Hermitian semigroup on the Hilbert spaces (\ref{s.12}-\ref{s.15}).  
The generator is a positive self-adjoint operator \cite{riesz}\cite{simon}.

Boosts present additional complications.  Even an infinitesimal
rotation in a Euclidean space time plane will map a general function
with positive Euclidean time support to one that violates this
condition.  The self-adjointness of the boost generator cannot be
demonstrated by showing that it defines a unitary one-parameter group
or contractive semigroup, however it turns out that rotations in
Euclidean space time planes, which are interpreted as boosts with
complex rapidity, define local symmetric semigroups
\cite{Klein:1981}\cite{Klein:1983} \cite{Frohlich:1983kp} on the
Hilbert spaces (\ref{s.12}-\ref{s.15}).  These have self-adjoint
generators, which are exactly the boost generators.

The conditions for a local symmetric semigroup 
\cite{Klein:1981} 
are
\begin{itemize}
\item[1.] For each $\theta \in [0 ,\theta_0]$, there is a linear subset
${\cal D}_{\theta}$ such that  ${\cal D}_{\theta_1}\supset  {\cal D}_{\theta_2}$
if $\theta_1< \theta_2$, and $\cup_{0<\theta<\theta_0}
{\cal D}_{\theta_2}$ is dense.

\item[2.] For each $\theta \in [0,\theta_0]$, $E(\theta)$ is a linear operator
  on the Hilbert space with domain ${\cal D}_{\theta}$

\item[3.] $E(0)=I$,
  $E(\theta_1):{\cal D}_{\theta_2}\to  {\cal D}_{\theta_2-\theta_1}$,
  and $E(\theta_1)E(\theta_2)= E(\theta_1+\theta_2)$ on
  ${\cal D}_{\theta_1+\theta_2}$ for $\theta_1,\theta_2,\theta_1+\theta_2 \in
  [0,\theta_0]$
  
\item[4.] $E(\theta)$ is Hermitian for $\theta \in [0,\theta_0]$ 
  
\item[5.] $E(\theta)$ is weakly continuous on $[0,\theta_0]$  

\end{itemize}

When these conditions are satisfied there is a unique self-adjoint
operator $K$ such that ${\cal D}_{\theta} \subset   {\cal D}_{e^{-K\theta}}$
and $E(\theta)$ is the restriction of $e^{-K\theta}$ to
${\cal D}_{\theta}$.

In this case $E(\theta)$ represents Euclidean space time rotations
considered as operators on the Hilbert space (\ref{a.4}) 
restricted to domains that will be described below.

The domains are Schwartz functions with space Euclidean time support the
wedge shaped region defined by
\beq
\mathbf{x} \cdot \hat{\mathbf{n}} - {x^0_e \over \epsilon} + \epsilon <0
\label{sa.5}
\eeq
\beq
\mathbf{x} \cdot \hat{\mathbf{n}} + {x^0_e \over \epsilon} - \epsilon >0
\label{sa.6}
\eeq

The wedge shaped region becomes the positive Euclidean time
half plane in the limit
that $\epsilon\to 0$.  Schwartz functions with support on this
half plane are dense.  In addition,
if this domain is rotated  by an angle less than 
$\theta_\epsilon :=\pm \tan^{-1}(\epsilon)$, it
will still be contained in the positive Euclidean time
half plane.  Schwartz functions with support in
these wedge shaped regions can be constructed from
Schwartz functions that have support or positive Euclidean time
by multiplying the function by 
$g(x^0,\mathbf{x}\cdot \hat{\mathbf{n}},\epsilon)$
where 
\beq
g(x^0,\mathbf{x}\cdot \hat{\mathbf{n}},\epsilon)) =
h({x^0_e \over \epsilon} - \epsilon +\mathbf{x} \cdot \hat{\mathbf{n}})
h({x^0_e \over \epsilon} - \epsilon -\mathbf{x} \cdot \hat{\mathbf{n}}).
\label{sa.7}
\eeq
and
\beq
h (\lambda)  =
\left \{
\begin{array}{cc}
e^{- {1 \over (\lambda)^2}} & \lambda >0 \\
0 & \lambda \leq 0 \\  
\end{array} 
\right . .
\label{sa.8}
\eeq
is a  smoothed Heaviside function.
$g(x^0,\mathbf{x}\cdot \hat{\mathbf{n}},\epsilon))$ is a Schwartz function with
support in the wedge shaped region (\ref{sa.5}-\ref{sa.6}) that
approaches $1$ as $\epsilon (\theta)$ approaches 0.

The domain ${\cal D}_{\theta}$ is taken as the space of Schwartz functions with
positive time support multiplied by the function
$g(x^0,\mathbf{x}\cdot \hat{\mathbf{n}},\epsilon))$ where $\theta=\theta_\epsilon$.  The Euclidean space time rotations restricted to
these  domains have all of the properties of local symmetric semigroup.
It follows that the boost generators $\mathbf{K}$ are self-adjoint
on the physical Hilbert space.

\section{Finite transformations}

Finite Poincar\'e transformations are needed for most applications.
While the generators for space translations and rotations were
constructed from the associated unitary one-parameter groups, the
construction of the boost generators and Hamiltonian was not as
direct. Because self adjointness was established for the Hamiltonian
and boost generators, each one of these generators has a dense set of
analytic vectors \cite{simon1} where exponential series for the unitary
one parameter groups converge.  This ensures that the differential
operators (\ref{d.27}) that define the boost generators
applied to a dense set of functions with positive time support have positive
time support.

Directly summing the exponential series is inefficient.  On the other
hand, the structure of the finite unitary transformations is fixed by
(\ref{c.29}) when they act on irreducible basis states.  The situation
is analogous to non-relativistic quantum mechanics - time evolution
becomes trivial once the Hamiltonian is diagonalized.  In the
Euclidean framework, the analogous problem is to diagonalize the mass
squared operator.  This is a dynamical problem that depends on the
choice of reflection positive Euclidean covariant distributions. For
the Green functions discussed in sections 4-5 the mass operator is the
four dimensional Euclidean Laplacian.  For more general Euclidean
covariant Green's functions is it second order differential operator.
The spectral condition ensures that there are no negative energy
states.

The Euclidean Green's functions are manifestly covariant
with respect to space translations and rotations.  Given a mass
eigenstate, the translational and rotational covariance can
be used to decompose the mass eigenstate into a
linear superposition of simultaneous eigenstates of linear momentum,
and spins.  On these states the unitary representation of the
Poincar\'e group acts irreducibly.

In the Euclidean formalism, since the dynamics
is in the Green function, mass eigenstates are solutions to 
\beq
\langle \phi \vert (M^2-m^2) \vert \psi \rangle =0  
\label{ft.1}
\eeq
for all $\phi$ satisfying the support condition.
Methods for constructing mass eigenstates satisfying the support
condition are discussed in \cite{Aiello:2015jgc}.

Mass-momentum eigenstates can be constructed using 
\beq
\vert m,\mathbf{p} \rangle := 
\int e^{-i \mathbf{p}\cdot \mathbf{a}}U(\mathbf{a}) \vert \psi \rangle d\mathbf{a}. 
\label{ft.2}
\eeq
Applying the translation operator $U(\mathbf{a}')$ to this vector gives
\[
U(\mathbf{a}')\vert m,\mathbf{p} \rangle  =
U(\mathbf{a}') \int e^{-i \mathbf{p}\cdot \mathbf{a}}U(\mathbf{a}) \vert \psi \rangle d\mathbf{a} =
\int e^{-i \mathbf{p}\cdot \mathbf{a}}U(\mathbf{a}+\mathbf{a}')
\vert \psi \rangle d\mathbf{a}=
\]
\beq
\int e^{-i \mathbf{p}\cdot (\mathbf{a}''-\mathbf{a}')}U(\mathbf{a}'')
\vert \psi \rangle d\mathbf{a}''=
e^{i \mathbf{p}\cdot \mathbf{a}'}
\int e^{-i \mathbf{p}\cdot \mathbf{a}''}U(\mathbf{a}'')
\vert \psi \rangle d\mathbf{a}'' =
e^{i\mathbf{p}\cdot \mathbf{a}'}
\vert m,\mathbf{p} \rangle
\label{ft.3}
\eeq
which shows that (\ref{ft.2}) is either 0 or an eigenstate of
linear momentum with eigenvalue $\mathbf{p}$.

The mass-momentum eigenstates can be decomposed into spin eigenstates.
Right and left handed kernels with the covariance properties
(\ref{s.12}) or (\ref{s.15}) after integration become kernels for the
covariant representations (\ref{d.13}) and (\ref{d.14}).  For
Green's functions with these rotational covariance properties
the covariant basis states, up to normalization, can be constructed
as follows,
\beq
\vert (m,s) \mathbf{p},\mu \rangle :=
\int U(R) \vert m,  R^{-1}\mathbf{p} \rangle D^{s*}_{\mu 0} [R] dR
\label{ft.4}
\eeq
where the integral is over the $SU(2)$ Haar measure.
For a fixed rotation $R'$:
\[
U(R') \vert (m,s) R^{\prime -1}\mathbf{p},\mu \rangle =
U(R')\int U(R) \vert m,  R^{-1} R^{\prime -1} \mathbf{p} \rangle D^{s*}_{\mu 0} [R] dR =
\int U(R' R) \vert m,  (R'R)^{-1}\mathbf{p}  \rangle D^{s*}_{\mu 0 } [R] dR =
\]
\beq
\int U(R'') \vert m,  R^{\prime\prime -1}\mathbf{p} \rangle D^{s*}_{\mu \nu} [R^{'-1}]
D^{s*}_{\nu 0}[R''] dR'' = 
\int U(R'') \vert m,  R^{\prime \prime -1}R' \mathbf{p} \rangle 
D^{s*}_{\nu 0}[R''] dR''D^{s}_{\nu \mu} [R'] =
\vert (m,s) \mathbf{p},\nu \rangle D^{s}_{\nu \mu} [R'].
\label{ft.5}
\eeq
If $R'$ is a rotation about the $z$ axis, it follows
that the resulting vector is an eigenstate of
$s^2$ and $s_z$.
This shows how mass eigenstates can be
decomposed into a superposition of Lorentz covariant states that
transform irreducibly with respect to the Poincar\'e group.

\section{Summary and Conclusion} 

The purpose of this paper is to provide explicit representations for
Poincar\'e generators for systems of particles of any spin in
Euclidean representations of relativistic quantum mechanics,
demonstrate that these generators satisfy the commutation relations of
the Poincar\'e Lie Algebra and are self-adjoint with respect to a
reflection positive scalar product.  This was done by starting with
irreducible unitary representations of the Poincar\'e group and
expressing them in a manifestly Lorentz covariant form.  The inner
product in the Lorentz covariant representation necessarily had a
non-trivial kernel, which could be expressed in terms of reflection
positive Green functions.  Expressions for the generators for any spin
were derived based on these relations.

While the results are specifically for positive mass irreducible
representations, they apply more generally since any unitary
representation of the Poincar\'e group can be decomposed into a direct
integral of positive-mass positive-energy irreducible representations.

Two consequences of the Osterwalder-Schrader reconstruction theorem
are (1) the locality axiom is logically independent of the other
Euclidean axioms and (2) the Hilbert space representation of the
quantum theory does not require explicit analytic continuation.  These
observations suggest the possibility of formulating phenomenological
non-local relativistic quantum mechanical models in a purely Euclidean
representation
\cite{Kopp:2011vv}\cite{Polyzou:2013nga}\cite{Aiello:2015jgc}.  The
new feature is that the dynamics appears in model Euclidean Green's
functions rather than in the Hamiltonian, which is a simple
differential operator.  One of the advantages of the Euclidean
formulation is that Euclidean Green's functions are moments of a
Euclidean path integral, which provides a formal connection to the
dynamics of Lagrangian field theories.  Models can be formulated by
perturbing products of free Green Euclidean functions with Euclidean
covariant interactions that preserve reflection positivity.

The authors would like to acknowledge Palle J{\o}rgensen for helpful discussions
on reflection positivity.

\bibliography{master_bibfile}

\begin{thebibliography}{34}
\expandafter\ifx\csname natexlab\endcsname\relax\def\natexlab#1{#1}\fi
\expandafter\ifx\csname bibnamefont\endcsname\relax
  \def\bibnamefont#1{#1}\fi
\expandafter\ifx\csname bibfnamefont\endcsname\relax
  \def\bibfnamefont#1{#1}\fi
\expandafter\ifx\csname citenamefont\endcsname\relax
  \def\citenamefont#1{#1}\fi
\expandafter\ifx\csname url\endcsname\relax
  \def\url#1{\texttt{#1}}\fi
\expandafter\ifx\csname urlprefix\endcsname\relax\def\urlprefix{URL }\fi
\providecommand{\bibinfo}[2]{#2}
\providecommand{\eprint}[2][]{\url{#2}}

\bibitem[{\citenamefont{Wigner}(1939)}]{Wigner:1939cj}
\bibinfo{author}{\bibfnamefont{E.~P.} \bibnamefont{Wigner}},
  \bibinfo{journal}{Annals Math.} \textbf{\bibinfo{volume}{40}},
  \bibinfo{pages}{149} (\bibinfo{year}{1939}).

\bibitem[{\citenamefont{Schwinger}(1958)}]{Schwinger:pna}
\bibinfo{author}{\bibfnamefont{J.~S.} \bibnamefont{Schwinger}},
  \bibinfo{journal}{Proc. Natl. Acad. Sci. U. S.}
  \textbf{\bibinfo{volume}{44}}, \bibinfo{pages}{956} (\bibinfo{year}{1958}).

\bibitem[{\citenamefont{Schwinger}(1959)}]{Schwinger:1959zz}
\bibinfo{author}{\bibfnamefont{J.}~\bibnamefont{Schwinger}},
  \bibinfo{journal}{Phys. Rev.} \textbf{\bibinfo{volume}{115}},
  \bibinfo{pages}{721} (\bibinfo{year}{1959}).

\bibitem[{\citenamefont{Streater and Wightman}(1980)}]{Wightman:1980}
\bibinfo{author}{\bibfnamefont{R.~F.} \bibnamefont{Streater}} \bibnamefont{and}
  \bibinfo{author}{\bibfnamefont{A.~S.} \bibnamefont{Wightman}},
  \emph{\bibinfo{title}{PCT, Spin and Statistics, and All That}}
  (\bibinfo{publisher}{Princeton Landmarks in Physics}, \bibinfo{year}{1980}).

\bibitem[{\citenamefont{Jost}(1965)}]{jost}
\bibinfo{author}{\bibfnamefont{R.}~\bibnamefont{Jost}},
  \emph{\bibinfo{title}{The General Theory of Quantized Fields}}
  (\bibinfo{publisher}{AMS}, \bibinfo{year}{1965}).

\bibitem[{\citenamefont{Symanzik}(1966)}]{Symanzik:1966}
\bibinfo{author}{\bibfnamefont{K.}~\bibnamefont{Symanzik}},
  \bibinfo{journal}{J. Math. Phys.} \textbf{\bibinfo{volume}{7}},
  \bibinfo{pages}{510} (\bibinfo{year}{1966}).

\bibitem[{\citenamefont{Symanzik}(1968)}]{Symanzik:1968zz}
\bibinfo{author}{\bibfnamefont{K.}~\bibnamefont{Symanzik}},
  \bibinfo{journal}{Conf. Proc.} \textbf{\bibinfo{volume}{C680812}},
  \bibinfo{pages}{152} (\bibinfo{year}{1968}).

\bibitem[{\citenamefont{Nelson}(1973)}]{Nelson:1973}
\bibinfo{author}{\bibfnamefont{E.}~\bibnamefont{Nelson}}, \bibinfo{journal}{J.
  Funct. Anal.} \textbf{\bibinfo{volume}{12}}, \bibinfo{pages}{97}
  (\bibinfo{year}{1973}).

\bibitem[{\citenamefont{Osterwalder and Schrader}(1973)}]{Osterwalder:1973dx}
\bibinfo{author}{\bibfnamefont{K.}~\bibnamefont{Osterwalder}} \bibnamefont{and}
  \bibinfo{author}{\bibfnamefont{R.}~\bibnamefont{Schrader}},
  \bibinfo{journal}{Commun. Math. Phys.} \textbf{\bibinfo{volume}{31}},
  \bibinfo{pages}{83} (\bibinfo{year}{1973}).

\bibitem[{\citenamefont{Osterwalder and Schrader}(1975)}]{Osterwalder:1974tc}
\bibinfo{author}{\bibfnamefont{K.}~\bibnamefont{Osterwalder}} \bibnamefont{and}
  \bibinfo{author}{\bibfnamefont{R.}~\bibnamefont{Schrader}},
  \bibinfo{journal}{Commun. Math. Phys.} \textbf{\bibinfo{volume}{42}},
  \bibinfo{pages}{281} (\bibinfo{year}{1975}).

\bibitem[{\citenamefont{Jorgensen and Olafsson}(1998)}]{palle}
\bibinfo{author}{\bibfnamefont{P.}~\bibnamefont{Jorgensen}} \bibnamefont{and}
  \bibinfo{author}{\bibfnamefont{G.}~\bibnamefont{Olafsson}},
  \bibinfo{journal}{Journal of Functional Analysis}
  \textbf{\bibinfo{volume}{158}}, \bibinfo{pages}{26} (\bibinfo{year}{1998}).

\bibitem[{\citenamefont{Neeb and O\'afsson}(2018)}]{olafsson}
\bibinfo{author}{\bibfnamefont{K.~H.} \bibnamefont{Neeb}} \bibnamefont{and}
  \bibinfo{author}{\bibfnamefont{G.}~\bibnamefont{O\'afsson}},
  \emph{\bibinfo{title}{Reflection Positivity, A Representation Theoretic
  Perspective}}, vol.~\bibinfo{volume}{32} (\bibinfo{publisher}{Springer, Cham,
  Switzerland}, \bibinfo{year}{2018}).

\bibitem[{\citenamefont{Jaffe}(2018)}]{Jaffe:2018ftu}
\bibinfo{author}{\bibfnamefont{A.}~\bibnamefont{Jaffe}} (\bibinfo{year}{2018}),
  \eprint{1802.07880}.

\bibitem[{\citenamefont{Sokolov}(1977)}]{Sokolov:1977}
\bibinfo{author}{\bibfnamefont{S.~N.} \bibnamefont{Sokolov}},
  \bibinfo{journal}{Dokl. Akad. Nauk SSSR} \textbf{\bibinfo{volume}{233}},
  \bibinfo{pages}{575} (\bibinfo{year}{1977}).

\bibitem[{\citenamefont{Coester and Polyzou}(1982)}]{Coester:1982vt}
\bibinfo{author}{\bibfnamefont{F.}~\bibnamefont{Coester}} \bibnamefont{and}
  \bibinfo{author}{\bibfnamefont{W.~N.} \bibnamefont{Polyzou}},
  \bibinfo{journal}{Phys. Rev.} \textbf{\bibinfo{volume}{D26}},
  \bibinfo{pages}{1348} (\bibinfo{year}{1982}).

\bibitem[{\citenamefont{Keister and Polyzou}(1991)}]{Keister:1991sb}
\bibinfo{author}{\bibfnamefont{B.~D.} \bibnamefont{Keister}} \bibnamefont{and}
  \bibinfo{author}{\bibfnamefont{W.~N.} \bibnamefont{Polyzou}},
  \bibinfo{journal}{Adv. Nucl. Phys.} \textbf{\bibinfo{volume}{20}},
  \bibinfo{pages}{225} (\bibinfo{year}{1991}).

\bibitem[{\citenamefont{Widder}(1931)}]{Widder:1931}
\bibinfo{author}{\bibfnamefont{D.~V.} \bibnamefont{Widder}},
  \bibinfo{journal}{Trans. Amer. Math. Soc.} \textbf{\bibinfo{volume}{33}},
  \bibinfo{pages}{851} (\bibinfo{year}{1931}).

\bibitem[{\citenamefont{Widder}(1934)}]{Widder:1934}
\bibinfo{author}{\bibfnamefont{D.~V.} \bibnamefont{Widder}},
  \bibinfo{journal}{Bull. Amer. Math. Soc.} \textbf{\bibinfo{volume}{40}},
  \bibinfo{pages}{321} (\bibinfo{year}{1934}).

\bibitem[{\citenamefont{Widder}(1941)}]{Widder:1941}
\bibinfo{author}{\bibfnamefont{D.~V.} \bibnamefont{Widder}},
  \emph{\bibinfo{title}{The Laplace Transform}} (\bibinfo{publisher}{Dover},
  \bibinfo{year}{1941}).

\bibitem[{\citenamefont{Polyzou}(2019)}]{Polyzou:2019a}
\bibinfo{author}{\bibfnamefont{W.~N.} \bibnamefont{Polyzou}},
  \bibinfo{journal}{Phys. Rev.} \textbf{\bibinfo{volume}{C99}},
  \bibinfo{pages}{025202} (\bibinfo{year}{2019}), \eprint{1809.09717}.

\bibitem[{\citenamefont{Rose}(1957)}]{rose}
\bibinfo{author}{\bibfnamefont{M.}~\bibnamefont{Rose}},
  \emph{\bibinfo{title}{Elementary Theory of Angular Momentum}}
  (\bibinfo{publisher}{Wiley}, \bibinfo{year}{1957}).

\bibitem[{\citenamefont{Bogoliubov and Shirkov}(1959)}]{bogoliubov}
\bibinfo{author}{\bibfnamefont{N.~N.} \bibnamefont{Bogoliubov}}
  \bibnamefont{and} \bibinfo{author}{\bibfnamefont{D.~V.}
  \bibnamefont{Shirkov}}, \emph{\bibinfo{title}{Introduction to the theory of
  quantized fields}} (\bibinfo{publisher}{Wiley-Interscience},
  \bibinfo{year}{1959}).

\bibitem[{\citenamefont{Wightman}(1960)}]{Wightman}
\bibinfo{author}{\bibfnamefont{A.~S.} \bibnamefont{Wightman}},
  \emph{\bibinfo{title}{{L'Invariance Dans La Mecanique Quantique
  Relativiste}}}, vol.~\bibinfo{volume}{7} (\bibinfo{publisher}{Hermann,
  Paris}, \bibinfo{year}{1960}).

\bibitem[{\citenamefont{Berestetskii et~al.}(1982)\citenamefont{Berestetskii,
  Lifshitz, and Pitaevskii}}]{berestetskii}
\bibinfo{author}{\bibfnamefont{V.~B.} \bibnamefont{Berestetskii}},
  \bibinfo{author}{\bibfnamefont{E.~M.} \bibnamefont{Lifshitz}},
  \bibnamefont{and} \bibinfo{author}{\bibfnamefont{L.~P.}
  \bibnamefont{Pitaevskii}}, \emph{\bibinfo{title}{Quantum Electrodynamics}}
  (\bibinfo{publisher}{Pergammon Press, Elmsford N.Y.}, \bibinfo{year}{1982}).

\bibitem[{\citenamefont{Riesz and Sz.Nagy}(1972)}]{riesz}
\bibinfo{author}{\bibfnamefont{F.}~\bibnamefont{Riesz}} \bibnamefont{and}
  \bibinfo{author}{\bibfnamefont{B.}~\bibnamefont{Sz.Nagy}},
  \emph{\bibinfo{title}{Functional Analysis}} (\bibinfo{publisher}{Ungar,
  N.Y.}, \bibinfo{year}{1972}).

\bibitem[{\citenamefont{Gilmm and Jaffe}(1981)}]{glimm}
\bibinfo{author}{\bibfnamefont{J.}~\bibnamefont{Gilmm}} \bibnamefont{and}
  \bibinfo{author}{\bibfnamefont{A.}~\bibnamefont{Jaffe}},
  \emph{\bibinfo{title}{Quantum Physics - A functional Integral Point of View}}
  (\bibinfo{publisher}{Springer}, \bibinfo{year}{1981}).

\bibitem[{\citenamefont{Reed and Simon}(1979)}]{simon}
\bibinfo{author}{\bibfnamefont{M.}~\bibnamefont{Reed}} \bibnamefont{and}
  \bibinfo{author}{\bibfnamefont{B.}~\bibnamefont{Simon}},
  \emph{\bibinfo{title}{Methods of Modern mathematical Physics}}, vol.
  \bibinfo{volume}{III Scattering Theory} (\bibinfo{publisher}{Academic Press},
  \bibinfo{year}{1979}).

\bibitem[{\citenamefont{Klein and L.}(1981)}]{Klein:1981}
\bibinfo{author}{\bibfnamefont{A.}~\bibnamefont{Klein}} \bibnamefont{and}
  \bibinfo{author}{\bibfnamefont{L.}~\bibnamefont{L.}}, \bibinfo{journal}{J.
  Functional Anal.} \textbf{\bibinfo{volume}{44}}, \bibinfo{pages}{121}
  (\bibinfo{year}{1981}).

\bibitem[{\citenamefont{Klein and L.}(1983)}]{Klein:1983}
\bibinfo{author}{\bibfnamefont{A.}~\bibnamefont{Klein}} \bibnamefont{and}
  \bibinfo{author}{\bibfnamefont{L.}~\bibnamefont{L.}}, \bibinfo{journal}{Comm.
  Math. Phys} \textbf{\bibinfo{volume}{87}}, \bibinfo{pages}{469}
  (\bibinfo{year}{1983}).

\bibitem[{\citenamefont{Frohlich et~al.}(1983)\citenamefont{Frohlich,
  Osterwalder, and Seiler}}]{Frohlich:1983kp}
\bibinfo{author}{\bibfnamefont{J.}~\bibnamefont{Frohlich}},
  \bibinfo{author}{\bibfnamefont{K.}~\bibnamefont{Osterwalder}},
  \bibnamefont{and} \bibinfo{author}{\bibfnamefont{E.}~\bibnamefont{Seiler}},
  \bibinfo{journal}{Annals Math.} \textbf{\bibinfo{volume}{118}},
  \bibinfo{pages}{461} (\bibinfo{year}{1983}).

\bibitem[{\citenamefont{Reed and Simon}(1972)}]{simon1}
\bibinfo{author}{\bibfnamefont{M.}~\bibnamefont{Reed}} \bibnamefont{and}
  \bibinfo{author}{\bibfnamefont{B.}~\bibnamefont{Simon}},
  \emph{\bibinfo{title}{Methods of Modern mathematical Physics}}, vol.
  \bibinfo{volume}{Functional Analysis} (\bibinfo{publisher}{Academic Press,
  N.Y.}, \bibinfo{year}{1972}).

\bibitem[{\citenamefont{Aiello and Polyzou}(2016)}]{Aiello:2015jgc}
\bibinfo{author}{\bibfnamefont{G.}~\bibnamefont{Aiello}} \bibnamefont{and}
  \bibinfo{author}{\bibfnamefont{W.}~\bibnamefont{Polyzou}},
  \bibinfo{journal}{Phys. Rev.} \textbf{\bibinfo{volume}{D93}},
  \bibinfo{pages}{056003} (\bibinfo{year}{2016}), \eprint{1512.03651}.

\bibitem[{\citenamefont{Kopp and Polyzou}(2012)}]{Kopp:2011vv}
\bibinfo{author}{\bibfnamefont{P.}~\bibnamefont{Kopp}} \bibnamefont{and}
  \bibinfo{author}{\bibfnamefont{W.}~\bibnamefont{Polyzou}},
  \bibinfo{journal}{Phys. Rev.} \textbf{\bibinfo{volume}{D85}},
  \bibinfo{pages}{016004} (\bibinfo{year}{2012}), \eprint{1106.4086}.

\bibitem[{\citenamefont{Polyzou}(2014)}]{Polyzou:2013nga}
\bibinfo{author}{\bibfnamefont{W.~N.} \bibnamefont{Polyzou}},
  \bibinfo{journal}{Phys. Rev.} \textbf{\bibinfo{volume}{D89}},
  \bibinfo{pages}{076008} (\bibinfo{year}{2014}), \eprint{1312.3585}.

\end{thebibliography}
\end{document}